\documentclass[aps,prx,a4paper,notitlepage,reprint,superscriptaddress,longbibliography]{revtex4-1}

\usepackage{amsmath,amssymb,amsfonts} 
\usepackage{mathrsfs} 

\usepackage{mathtools} 
\usepackage{color}
\usepackage{chemformula}
\usepackage{siunitx}
\usepackage{graphicx,float}
\usepackage[colorlinks,
linkcolor=red,
citecolor=blue,
urlcolor=red]{hyperref}

\usepackage{cleveref}
\usepackage{soul}
\usepackage{changes}

\renewcommand{\t}[1]{\mathrm{#1}}
\newcommand{\SiN}{Si$_3$N$_4\,$}

\newcommand{\mbrsize}[1]{#1\,mm}

\newcommand{\figref}[1]{Fig.~\ref{#1}}
\renewcommand{\eqref}[1]{Eq.~\ref{#1}}
\newcommand{\secref}[1]{Sec.~\ref{#1}}

\begin{document}
\title{Thermal intermodulation noise in cavity-based measurements}
	
\author{S. A. Fedorov}
\thanks{These authors contributed equally}
\email{sergey.fedorov@epfl.ch}
\affiliation{Institute of Physics, Swiss Federal Institute of Technology Lausanne (EPFL), 1015 Lausanne, Switzerland}
	
\author{A. Beccari}
\thanks{These authors contributed equally}
\affiliation{Institute of Physics, Swiss Federal Institute of Technology Lausanne (EPFL), 1015 Lausanne, Switzerland}

\author{A. Arabmoheghi}
\affiliation{Institute of Physics, Swiss Federal Institute of Technology Lausanne (EPFL), 1015 Lausanne, Switzerland}

\author{D. J. Wilson}
\affiliation{College of Optical Sciences, University of Arizona, Tucson, Arizona 85721, USA}

\author{N. J. Engelsen}
\email{nils.engelsen@epfl.ch}
\affiliation{Institute of Physics, Swiss Federal Institute of Technology Lausanne (EPFL), 1015 Lausanne, Switzerland}

\author{T. J. Kippenberg}
\email{tobias.kippenberg@epfl.ch}
\affiliation{Institute of Physics, Swiss Federal Institute of Technology Lausanne (EPFL), 1015 Lausanne, Switzerland}
	
\begin{abstract}
Thermal frequency fluctuations in optical cavities limit the sensitivity of precision experiments ranging from gravitational wave observatories to optical atomic clocks.
Conventional modeling of these noises assumes a linear response of the optical field to the fluctuations of cavity frequency.
Fundamentally, however, this response is nonlinear.
Here we show that nonlinearly transduced thermal fluctuations of cavity frequency can dominate the broadband noise in photodetection, even when the magnitude of fluctuations is much smaller than the cavity linewidth.
We term this noise ``thermal intermodulation noise'' and show that for a resonant laser probe it manifests as intensity fluctuations.
We report and characterize thermal intermodulation noise in an optomechanical cavity, where the frequency fluctuations are caused by mechanical Brownian motion, and find excellent agreement with our developed theoretical model.
We demonstrate that the effect is particularly relevant to quantum optomechanics: using a phononic crystal $\ch{Si3N4}$ membrane with a low mass, soft-clamped mechanical mode we are able to operate in the regime where measurement quantum backaction contributes as much force noise as the thermal environment does.
However, in the presence of intermodulation noise, quantum signatures of measurement are not revealed in direct photodetection.
The reported noise mechanism, while studied for an optomechanical system, can exist in any optical cavity.

\end{abstract}
	
\date{\today}
\maketitle
	

\section{Introduction}
Optical cavities are an enabling technology for precision experiments, including gravitational wave detection \cite{ligo_collaboration_observation_gw_2016}, ultrastable lasers \cite{sterr_ultrastable_2009}, cavity quantum electrodynamics \cite{ye_quantum_2008} and cavity optomechanics \cite{aspelmeyer_cavity_2014}.
Depending on the application, they can be exquisite detectors for performing quantum-limited measurements \cite{clerk_introduction_2010}, or highly stable frequency references for clock lasers \cite{sterr_ultrastable_2009}.
In both cases, performance is limited by fundamental thermodynamic frequency fluctuations that exist in any cavity at finite temperature, as a result of the Brownian motion of mirror surfaces, fluctuations of the refractive index or thermoelastic effect \cite{braginsky_thermorefractive_2000,gorodetsky_thermal_noise_compensation_2008}.
Minimizing these fluctuations has played a key role in the design of interferometric gravitational wave detectors  \cite{braginsky_thermodynamical_1999}, motivated the development of crystalline mirror coatings \cite{cole_tenfold_2013}, and cryogenic operation of reference cavities for optical frequency metrology \cite{robinson_crystalline_2019}.

Thermal noises are particularly strong in optical cavities at the micro- and nanoscale. High-finesse optical microcavities are increasingly employed in a variety of precision measurements, ranging from compact reference cavities using crystalline microresonators \cite{matsko_whispering-gallery-mode_2007,alnis_thermal-noise-limited_2011,lee_spiral_2013} to photonic integrated microresonator based quantum optics experiments \cite{dutt_-chip_2015,hoff_integrated_2015,otterpohl_squeezed_2019} as well as optical frequency metrology, where driven Kerr nonlinear microresonators produce octave spanning combs \cite{kippenberg_dissipative_2018}, whose carrier envelope frequency is thermal noise limited \cite{drake_thermal_2019}.
In such microcavities thermal fluctuations are particularly prominent, because of the small mode volume \cite{gorodetsky_fundamental_2004,anetsberger_near-field_2009,huang_thermorefractive_2019}.

The conventional framework in which optical measurements are described assumes a linear transduction of cavity frequency fluctuations into the optical field, justified by the frequency excursions being small compared to the cavity linewidth. However, the nonlinearity of transduction is inherently present in any cavity and gives rise to qualitatively new phenomena: It results in the conversion of  Gaussian fluctuations of cavity frequency into non-Gaussian fluctuations of the optical field.
When a cavity is coupled to a quantum system, this
phenomenon been proposed for performing nonlinear quantum measurements \cite{vanner_selective_2011,brawley_nonlinear_2016,leijssen_nonlinear_2017}, which cannot be described within the leading order perturbation theory \cite{clerk_introduction_2010}.
At the same time the nonlinear conversion of thermal frequency fluctuations can impose qualitatively new constraints on a broad range of precision experiments, which to date have not been analyzed.

Here we report for the first time that the nonlinear modulation of the optical field by thermal frequency fluctuations can manifest as a broadband added noise in detection. We refer to this noise as \emph{thermal intermodulation noise} (TIN), since it mixes different Fourier components of cavity frequency fluctuations. This noise dominates when the linearly transduced thermal fluctuations are small, such as when detecting the intensity of near-resonant optical probe. As it is the leading-order contribution, TIN it is not necessarily negligible even when the nonlinearity of cavity transduction is small.

We experimentally observe and study TIN in a membrane-in-the-middle (MIM) optomechanical system \cite{thompson_strong_2008,wilson_cavity_2009}
---a promising platform for room temperature quantum optomechanical experiments \cite{cripe_measurement_2019,yap_broadband_2019}---and find excellent agreement with our developed theoretical model.
Using a \ch{Si3N4} membrane resonator hosting a high-$Q$ and low mass soft-clamped mode \cite{tsaturyan_ultracoherent_2017,reetz_analysis_2019}, we operate at a nominal quantum cooperativity of unity, i.e. in the regime where the linear measurement quantum backaction (arising from radiation pressure quantum fluctuations) is expected to overwhelm the thermal motion. 
This regime is required for a range of quantum enhanced measurement protocols \cite{vyatchanin_quantum_1995,sudhir_quantum_2017,kampel_improving_2017}, or generation of optical squeezed states \cite{safavi-naeini_squeezed_2013,purdy_strong_2013}.
Yet, the nonlinearity of our cavity prevents the observation of quantum correlations between the field quadratures, and manifests itself in TIN significantly above the shot noise (i.e. quantum noise) level.
Surprisingly, we find that TIN dominates the fluctuations of the intensity of the optical field even when the thermally induced frequency fluctuations are substantially smaller than the cavity linewidth.
Since TIN is a coherent effect, it only requires the knowledge of spectrum of cavity frequency fluctuations to be modeled, and our experimental data is well matched by a model with no free parameters.

We show that for a particular ``magic" detuning from the cavity TIN is fully cancelled in direct detection, and propose a more general cancellation scheme suitable for arbitrary detuning. Our observations, while made for an optomechanical system, are broadly applicable, irrespective of the underlying thermal noise source. Thermal intermodulation noise can be of relevance to any cavity based measurement schemes at finite temperature.

\section{Theory of thermal intermodulation noise}
\label{sec:genTheor}

We begin by presenting the theory of thermal intermodulation noise with the assumption that the cavity frequency fluctuations are slow compared to the optical decay rate. We concentrate on the lowest-order, i.e. quadratic, nonlinearity of the cavity detuning transduction.
We consider (as in our experimental setup) an optical cavity with two ports, which is driven by a laser coupled to port one. The output from port two is directly detected on a photodiode. In the classical regime, i.e. neglecting vacuum fluctuations, the complex amplitude of the intracavity optical field, $a$, and the output field $s_\t{out,2}$ can be found from the input-output relations
\begin{align}
&\frac{da(t)}{dt}=\left(i\Delta(t)-\frac{\kappa}{2}\right)a(t)+\sqrt{\kappa_1}\, s_\t{in,1},\\
& s_\t{out,2}(t)=-\sqrt{\kappa_2}a(t),\label{eq:oi2}
\end{align}
where $s_\t{in,1}$ is the constant coherent drive amplitude, $\Delta(t)=\omega_L-\omega_c(t)$ is the laser detuning from the cavity resonance, modulated by the cavity frequency noise, and $\kappa_{1,2}$ are the external coupling rates of ports one and two ($\kappa_1=\kappa_2$ in our case) and $\kappa=\kappa_1+\kappa_2$. In the fast cavity limit, when the optical field adiabatically follows $\Delta(t)$, the intracavity field is found as
\begin{equation}\label{eq:aFull}
a(t)=2\sqrt{\frac{\eta_1}{\kappa}} L(\nu(t))\, s_\t{in,1},
\end{equation}
where we introduced for brevity the normalized detuning $\nu=2\Delta/\kappa$, the cavity decay ratios $\eta_{1,2}=\kappa_{1,2}/\kappa$  and Lorentzian susceptibility
\begin{equation}
L(\nu)=\frac{1}{1-i\nu}.
\end{equation}
Expanding $L$ in \eqref{eq:aFull} over small detuning fluctuations $\delta\nu$ around the mean value $\nu_0$ up to second order we find the intracavity field as
\begin{equation}\label{eq:aSq}
a=2\sqrt{\frac{\eta_1}{\kappa}}L(\nu_0)(1+iL(\nu_0)\delta\nu -L(\nu_0)^2\delta\nu^2) s_\t{in,1}.
\end{equation}
According to \eqref{eq:aSq}, the intracavity field is modulated by the cavity frequency excursion, $\delta\nu$, and the frequency excursions squared, $\delta\nu^2$. If $\delta\nu(t)$ is a stationary Gaussian noise process, like typical thermal noises, the linear and quadratic contributions are uncorrelated (despite clearly not being independent). This is due to the fact that odd-order correlations vanish for Gaussian noise,
\begin{equation}
\langle\delta\nu(t)^2\delta\nu(t+\tau)\rangle=0,
\end{equation}
where $\langle ...\rangle$ is the time average, for an arbitrary time delay $\tau$.
Next, we consider the photodetected signal, which, up to a conversion factor, equals the intensity of the output light and is found to be
\begin{multline}\label{eq:detectedNoise}
I(t)=|s_\t{out,2}(t)|^2\propto \\
|L(\nu_0)|^2 \left(1-\frac{2\nu_0}{1+\nu_0^2}\delta\nu(t)+\frac{3\nu_0^2-1}{(1+\nu_0^2)^2}\delta\nu(t)^2\right).
\end{multline}
Notice that $\delta\nu(t)$ and $\delta\nu(t)^2$ can be distinguished by their detuning dependence. The linearly transduced fluctuations vanish on resonance ($\nu_0=0$), where $\partial L/\partial \nu=0$. Similarly, when $\partial^2 L/\partial\nu^2=0$, the quadratic frequency fluctuations vanish, and thus also the thermal intermodulation noise. We refer to the corresponding detuning values,
\begin{equation}
\nu_0=\pm 1/\sqrt{3},
\end{equation}
as ``magic''.
In the following experiments, we will make measurements at $\nu_0=-1/\sqrt{3}$ and $\nu_0=0$ to independently characterize the spectra of $\delta\nu(t)$ and $\delta\nu(t)^2$, respectively.

The total spectrum \cite{specta_notations} of the detected signal, $I(t)$, is an incoherent sum of the linear term given by,
\begin{equation}\label{eq:Snu}
S_{\nu\nu}[\omega]=\int_{-\infty}^{\infty}\langle \delta\nu(t) \delta\nu(t+\tau) \rangle e^{i\omega \tau}d\tau,
\end{equation}
and the quadratic term, which for Gaussian noise can be found using Wick's theorem \cite{gardiner_handbook_1985}
\begin{equation}
\langle \delta\nu(t)^2 \delta\nu(t+\tau)^2 \rangle=\langle\delta\nu(t)^2 \rangle^2+2\langle \delta\nu(t) \delta\nu(t+\tau) \rangle^2,
\end{equation}
as
\begin{multline}\label{eq:Snu2}
S_{\nu\nu}^{(2)}[\omega]=\int_{-\infty}^{\infty}\langle \delta\nu(t)^2 \delta\nu(t+\tau)^2 \rangle e^{i\omega \tau}d\tau=\\
2\pi\langle \delta\nu^2\rangle^2 \delta[\omega]+2\times\frac{1}{2\pi} \int_{-\infty}^{\infty} S_{\nu\nu}[\omega']S_{\nu\nu}[\omega-\omega']d\omega',
\end{multline}
where $\delta[\omega]$ is the Dirac delta function.

\begin{figure}[t]
\centering
\includegraphics[width=\columnwidth]{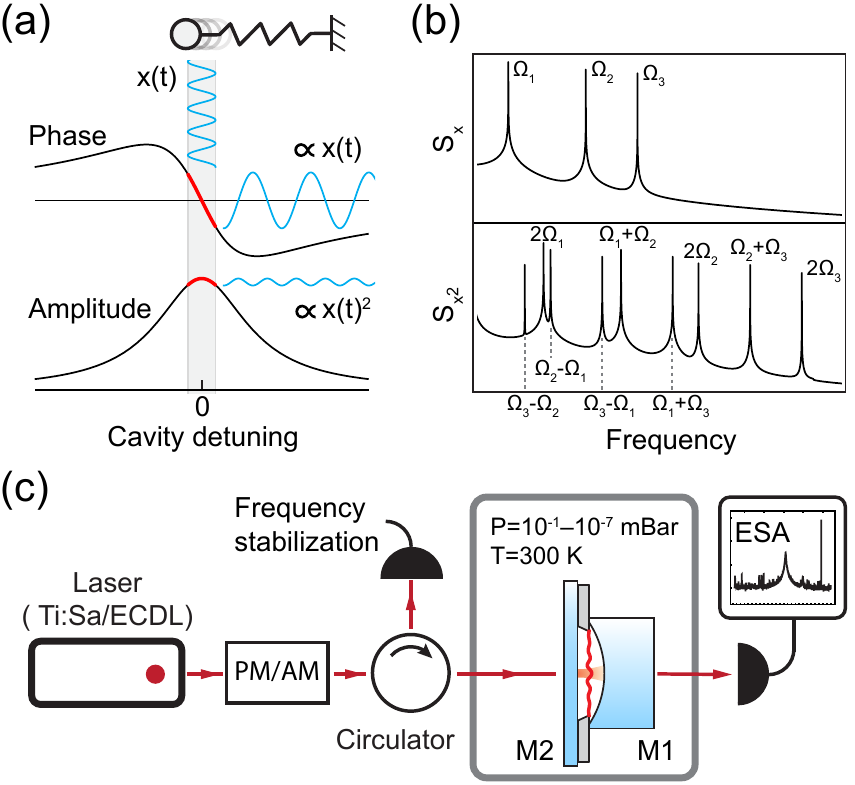}
\caption{Physical mechanism of optomechanical thermal intermodulation noise. \small{a) Transduction of the oscillator's motion to the phase (upper panel) and amplitude (lower panel) quadratures of resonant intracavity light. b) Spectra of linear (upper panel) and quadratic (lower panel) position fluctuations of a multimode resonator, showing the emergence of wideband noise. c) Experimental setup in which TIN is studied consisting of a membrane-in-the-middle optomechanical system. PM/AM: Phase/amplitude modulator. ESA: Electronic spectrum analyzer.}}
\label{fig:intro}
\end{figure}

\begin{figure*}[t]
\centering
\includegraphics[width=\textwidth]{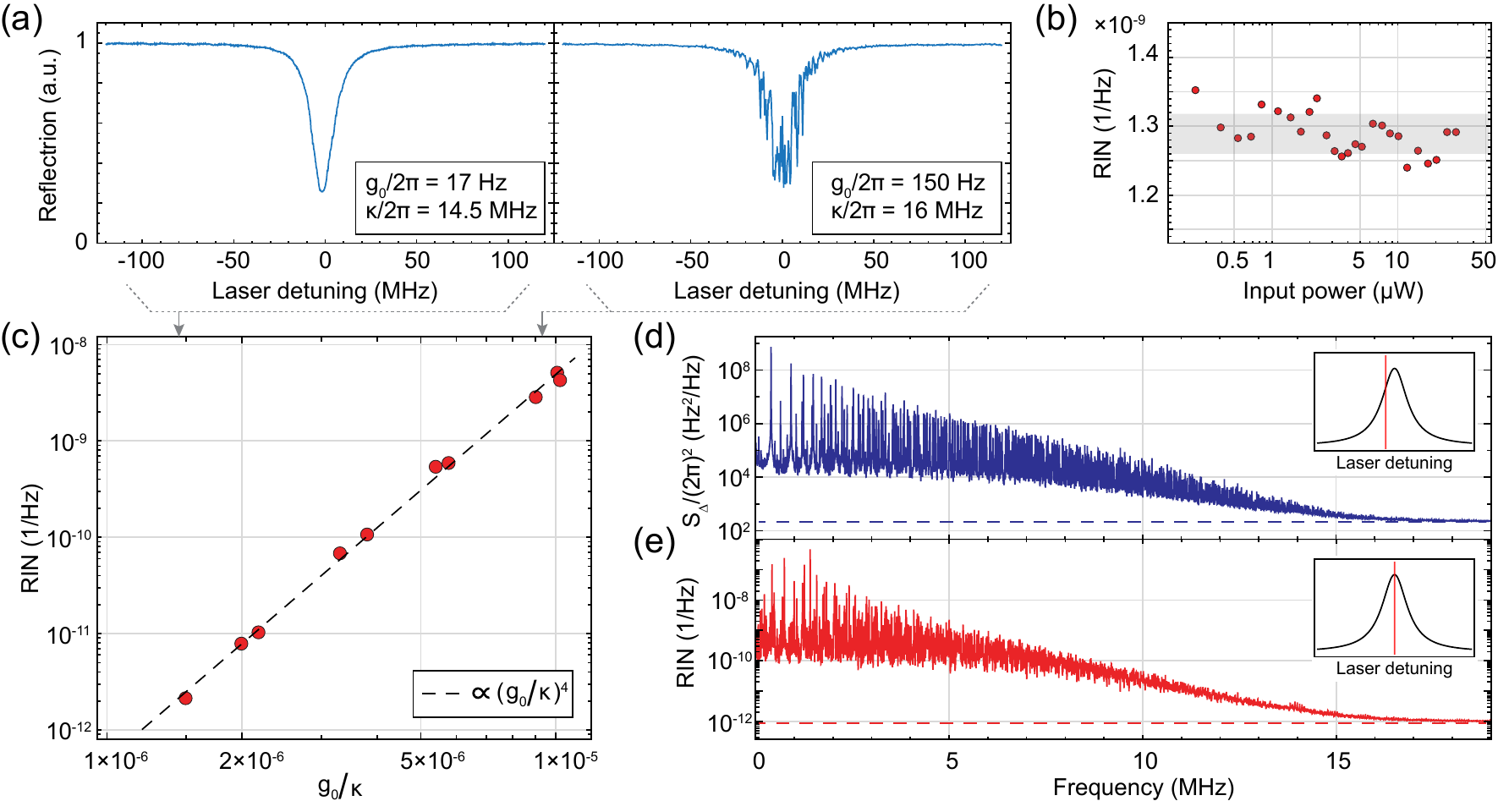}
\caption{Observation of optomechanical thermal intermodulation noise. \small{a), b) and c) show measurements for a membrane-in-the middle cavity with a 2 mm square membrane. a) Cavity reflection signal as the laser is scanned over two resonances, with low (left) and high (right) optomechanical coupling. b) Dependence of resonant RIN, averaged over $0.6-1.6$ MHz, on the input power. Parameters: $\kappa/2\pi=9.9$ MHz, $g_0/2\pi=84$ Hz for the fundamental mode. The interval of $\pm$ one standard deviation around the mean is shaded gray. c) Dependence of the average RIN in a $0.6-1.6$ MHz band on $g_0/\kappa$. d) Detuning noise of a MIM cavity with a 1 mm square membrane, $\kappa/2\pi=26.6$ MHz and $g_0/2\pi=330$ Hz for the fundamental mode, measured at the laser detuning $2\Delta/\kappa\approx -1/\sqrt{3}$. e) Resonant RIN measured under the same conditions as in (d) but at $\Delta=0$.}}
\label{fig:rectMbr}
\end{figure*}

\section{Brownian intermodulation noise}

In an optomechanical cavity, the dominant source of cavity frequency fluctuations is the Brownian motion of mechanical modes coupled to the cavity,
\begin{equation}\label{eq:dnuOpt}
\delta\nu(t)= 2\frac{G}{\kappa} x(t),
\end{equation}
where $G=-\partial \omega_c/\partial x$ is the linear optomechanical coupling constant, and $x$ is the total resonator displacement, i.e. the sum of independent contributions $x_n$ of different mechanical modes (the effect of the finite cavity mode waist is treated in Appendix~\ref{sec:theoryDetails}). The spectrum of the Brownian frequency noise is then found to be
\begin{equation}\label{eq:mbrFreqFluct}
S_{\nu\nu}[\omega]=\left(\frac{2G}{\kappa}\right)^{2}\sum_n S_{xx,n}[\omega],
\end{equation}
where $S_{xx,n}[\omega]$ are the displacement spectra of individual mechanical modes (see Appendix~\ref{sec:theoryDetails} for more details). The thermomechanical frequency noise given by \eqref{eq:mbrFreqFluct} produces TIN which contains peaks at sums and differences of mechanical resonance frequencies and a broadband background due to the off-resonant components of thermal noise, as illustrated in \figref{fig:intro}b. The magnitude of the intermodulation noise is related to the quadratic spectrum of the total mechanical displacement, $S_{xx}^{(2)}$, as
\begin{equation}\label{eq:intNoiseSxx2}
S_{\nu\nu}^{(2)}=(2G/\kappa)^{4}S_{xx}^{(2)}.
\end{equation}

A reservation needs to be made: the theory presented in \secref{sec:genTheor} is only strictly applicable to an optomechanical cavity when the input power is sufficiently low, such that the driving of mechanical motion by radiation pressure fluctuations created by the intermodulation noise is negligible; otherwise the fluctuations of $x(t)$ and $\delta \nu(t)$ may deviate from purely Gaussian and correlations exist between $\delta\nu(t)$ and $\delta\nu(t)^2$. On a practical level, this reservation has minor significance for our experiment. Also, the presence of linear dynamical backaction of radiation pressure does not change the results of \secref{sec:genTheor} but does modify $S_{xx}$.

Thermal intermodulation noise can preclude the observation of linear quantum correlations, which are induced by the vacuum fluctuations of radiation pressure between the quadratures of light and manifest as ponderomotive squeezing \cite{purdy_strong_2013,safavi-naeini_squeezed_2013}, Raman sideband asymmetry \cite{sudhir_appearance_2017} and the cancellation of shot noise in force measurements \cite{kampel_improving_2017,sudhir_quantum_2017}. The observation of quantum correlations typically requires selecting a mechanical mode with high quality factor, $Q$, a spectral neighbourhood free from other modes, and a high optomechanical coupling rate. If TIN is taken into account, simply increasing the quantum cooperativity is not sufficient, and the following condition also needs to be satisfied:
\begin{equation}\label{eq:qbaCond}
C_q\left(\frac{g_0}{\kappa}\right)^2 \Gamma_m \bar{n}_\t{th}\frac{S^{(2)}_{xx}[\omega]}{x_\t{zpf}^4}\ll 1.
\end{equation}
Here $C_q=4 g_0^2\bar{n}_c/(\kappa \Gamma_m \bar{n}_\t{th})\gtrsim 1$ is the quantum cooperativity, $g_0=G\sqrt{\hbar /(2 m_\t{eff} \Omega_m)}$,  $\bar{n}_c$ is the intracavity photon number. The selected mechanical mode is characterized by the resonance frequency $\Omega_m$, the damping rate $\Gamma_m$, the effective mass $m_\t{eff}$ and the thermal phonon occupancy $\bar{n}_\t{th}=\hbar\Omega_m/(k_B T)$.
From the condition given by \eqref{eq:qbaCond}, one can immediately observe that by reducing the mechanical dissipation and $g_0/\kappa$, one can keep the quantum cooperativity constant while lowering the intermodulation noise. The engineering of the mode spectrum to reduce $S^{(2)}_{xx}$ at the desired frequency might also be a fruitful approach.

The nonlinearity of the cavity-laser detuning response, which produces TIN, modulates the optical field proportional to $x^2$ in a way analogous to, but not equivalent to, quadratic optomechanical coupling, $\partial^2 \omega_c/\partial x^2$. It was noticed that the cavity transduction commonly results in effective quadratic coupling which is orders of magnitude stronger than the highest experimentally reported $\partial^2 \omega_c/\partial x^2$  (in terms of the optical signal proportional to $x^2$ \cite{brawley_nonlinear_2016}). In Appendix~\ref{sec:quadOptom}, it is shown that the same is true in the MIM system. Here, the quadratic signal originating from nonlinear transduction, which creates the intermodulation noise, is larger than the signals due to nonlinear optomechanical coupling, $\partial^2 \omega_c/\partial x^2$, by a factor of $r \mathcal{F}$, where $r$ is the membrane reflectivity and $\mathcal{F}$ is the optical finesse.

\section{Experimental observation of thermal intermodulation noise}

Our experimental setup, shown in \figref{fig:intro}c, comprises a membrane-in-the-middle cavity, consisting of two high-reflectivity mirrors and a chip with high-stress stoichiometric \SiN membrane sandwiched directly between them. The MIM cavity is situated in a vacuum chamber at room temperature and probed in transmission. See Appendix~\ref{sec:expDetails} for more details.

\subsection{Intermodulation noise in a cavity with a uniform membrane}
\label{sec:rectMbr}
We first characterize the TIN in cavities with 20 nm-thick uniform square membranes of different sizes. The optomechanical cooperativity was kept low in order to eliminate dynamical backaction of the light; achieved by increasing the vacuum pressure and keeping the mechanical modes gas damped to $Q\sim 10^3$. The reflection signals of two resonances of a MIM cavity with a \mbrsize{2} membrane are presented in \figref{fig:rectMbr}a. The resonances have similar optical linewidths (about 15 MHz) but their optomechanical coupling is different by a factor of ten. The resonance with high coupling ($g_0/2\pi=150$ Hz) shows clear signatures of thermal noise. For this resonance the total r.m.s. thermal frequency fluctuations are expected to be around 2 MHz, which is still well below the cavity linewidth, $\kappa/2\pi=16$ MHz.

Thermal fluctuations of the reflection signal are clearly observed in the right panel of \figref{fig:rectMbr}a even when the laser is resonant with the cavity. This is not expected in linear optomechanics, where the mechanical motion only modulates the phase of a resonant laser probe. Typical spectra of the detected noise are shown in \figref{fig:rectMbr}d for a cavity with a different, \mbrsize{1}, square membrane. With the laser detuned from the cavity resonance (close to the ``magic" detuning, $\nu_0\approx -1/\sqrt{3}$), the transmission signal is dominated by the Brownian motion of membrane modes transduced by the cavity (shown in \figref{fig:rectMbr}d), in agreement with the prediction of linear optomechanics. The magnitude of thermomechanical noise is gradually reduced at high frequencies due to the averaging of membrane mode profiles \cite{zhao_wilson_suppression_2012,wilson_thesis_2012} over the cavity waist, until it meets shot noise at around 15 MHz. With the laser on resonance, from linear optomechanics it is expected that the output signal is shot noise limited. However, the experimental signal (shown in \figref{fig:rectMbr}e) contains a large amount of thermal noise---at an input power of 5 $\mu$W the classical RIN exceeds the shot noise level by about 25 dB at MHz frequencies. The spectrum of the resonant RIN is different from the spectrum of detuning fluctuations, owing to the nonlinear origin of the noise. At high frequency, the RIN level approaches shot noise, as verified by the optical power dependence (see Appendix~\ref{sec:expDetails}).

An unambiguous proof of the intermodulation origin of the resonant intensity noise is obtained by examining the scaling of the noise level with $G/\kappa$. In thermal equilibrium, the spectral density of frequency fluctuations, $\delta\nu(t)$, created by a particular membrane is proportional to $(G/\kappa)^2$, and therefore the spectral density of intermodulation noise is expected to be proportional to $(G/\kappa)^4$. We confirm this scaling by measuring the resonant intensity noise for different optical resonances of a cavity with a \mbrsize{2} membrane and present in \figref{fig:rectMbr}b the average noise magnitude as a function of $g_0/\kappa$, where $g_0$ is the optomechanical coupling of the fundamental mode. By performing a sweep of the input laser power on one of the resonances of the same cavity we show (see \figref{fig:rectMbr}b) that the resonant intensity noise level is power-independent and therefore the noise is not related to radiation pressure effects.

\begin{figure}[t]
\centering
\includegraphics[width=\columnwidth]{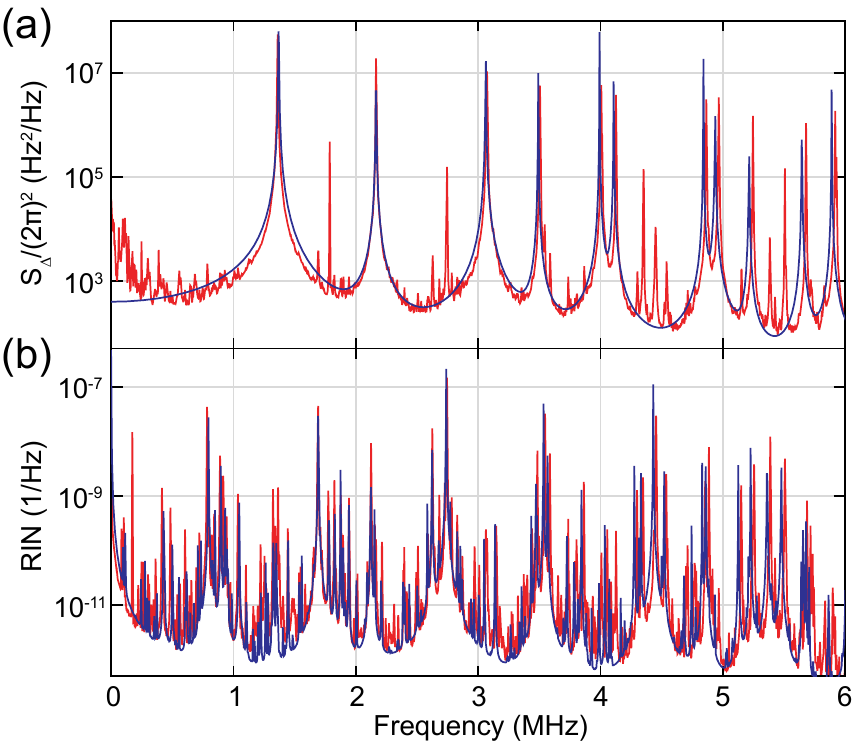}
\caption{\small{Comparison of theoretical and experimental frequency and resonant intensity noises. a) Detuning fluctuation and b) relative intensity noise spectra produced by the modes of a 20-nm thick, \mbrsize{0.3}, rectangular, Si$_3$N$_4$ membrane. Red shows experimental data and blue is the theoretical prediction.}}
\label{fig:theor}
\end{figure}

The TIN observed in our experiments agrees well with our theoretical model. By calculating the spectrum of total membrane fluctuations according to  \eqref{eq:mbrFreqFluct} and applying the convolution formula from \eqref{eq:Snu2} (see Appendix~\ref{sec:theoryDetails} for full details), we can accurately reproduce the observed noise. In \figref{fig:theor}, we compare the measured detuning and intensity noise spectra with the theoretical model. Here, we assume that the damping rates of all the membrane modes are identical, as the experiment is operated in the gas-damping-dominated regime. While this model is not detailed enough to reproduce all the noise features, it accurately reproduces the overall magnitude and the broadband envelope of the intermodulation noise observed in the experiment.

\begin{figure}[t]
\centering
\includegraphics[width=\columnwidth]{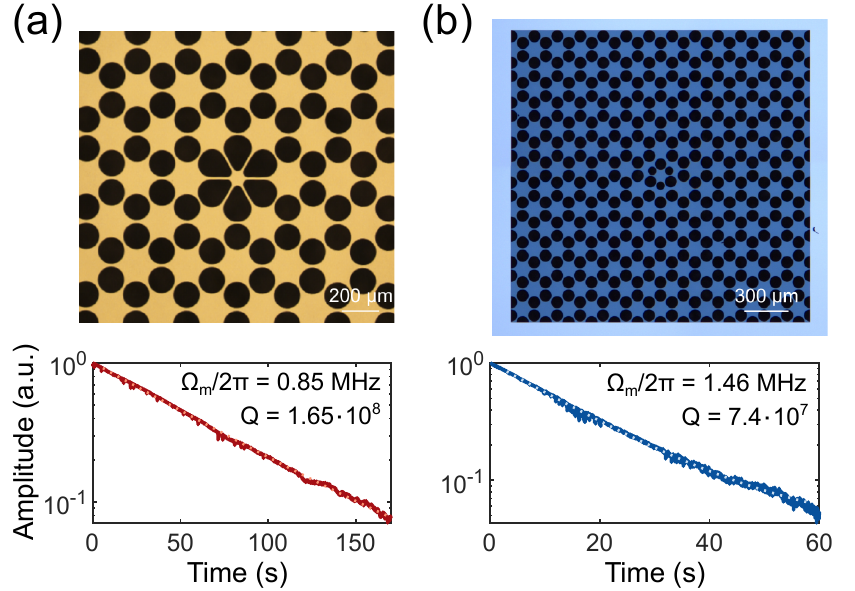}
\caption{\small{Microscope images of PnC membranes (top) and ringdowns of their soft-clamped, localized modes (bottom). a) 3.6\,mm$\times$3.3\,mm$\times$40\,nm, with a localized mode at 853 kHz, b) 2\,mm$\times$2\,mm$\times$20\,nm membrane with a localized mode at 1.46 MHz.}}
\label{fig:softClamped}
\end{figure}

\begin{figure*}[t]
\centering
\includegraphics[width=\textwidth]{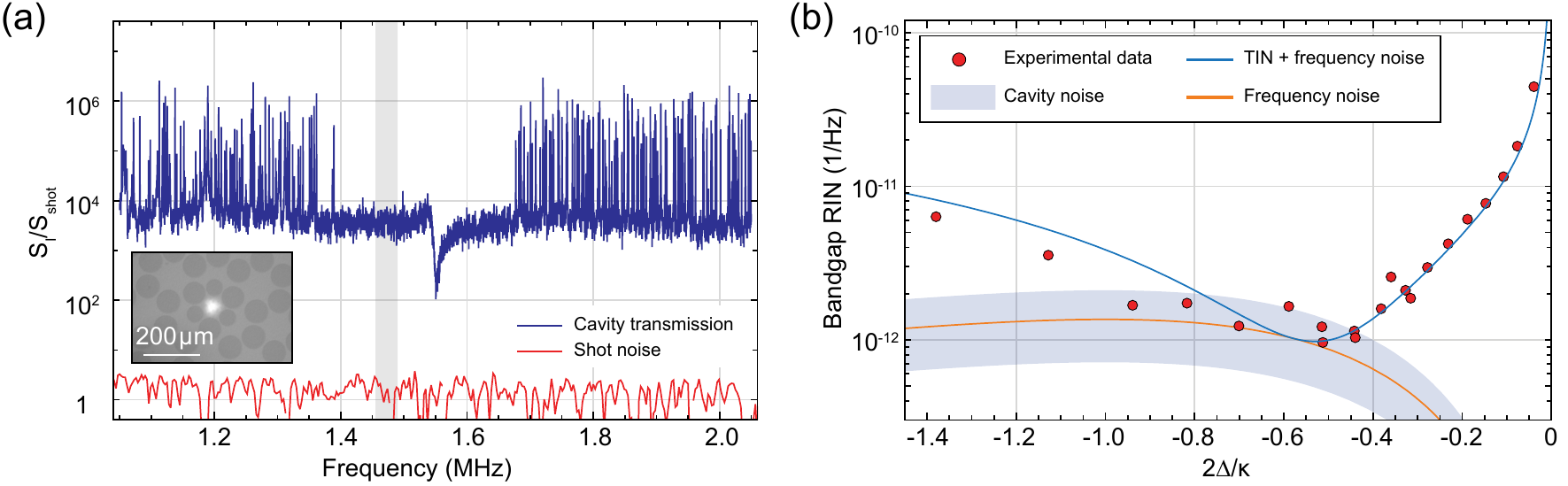}
\caption{\small{Measurement of the frequency spectrum and detuning dependence of thermal optomechanical intermodulation noise with a phononic crystal membrane.
a) Blue---photocurrent noise spectrum detected with the cavity-laser detuning set to $2\Delta/\kappa\approx-0.3$, red---shot noise level. The shaded region shows the noise averaging band for the plot in b. The inset shows an optical cavity mode (imaged at $\lambda\approx780$ nm) overlapping with the PnC membrane defect.  b) Variation of the relative intensity noise at bandgap frequencies with cavity-laser detuning. Red circles are experimental measurements, blue line---fit, orange line---cavity phase noise inferred from the fit, shaded blue region---independently calibrated cavity noise, with uncertainty from the selection of the averaging band (see Appendix~\ref{sec:expDetails}).}}
\label{fig:detuningSweep}
\end{figure*}

We would like to address two potential confounding effects: \emph{laser frequency noise} and  \emph{dissipative coupling}. The intensity noise of the laser was below $10^{-12}$ Hz$^{-1}$ for frequencies above $100$ kHz and therefore negligible in all resonant RIN measurements. In the same frequency range, the frequency noise of the laser is below 1 Hz$^2$/Hz, which is, again, much lower than the thermomechanical noise. For more details on the extraneous noises see Appendix~\ref{sec:expDetails}. As dissipative coupling leads to the modulation of optical linewidth by mechanical position, it could also potentially explain intensity noise in a resonant optical field. Although dissipative coupling is generally present in MIM cavities \cite{wilson_thesis_2012}, the magnitude of this noise is expected to be orders of magnitude below that measured in our experiments (see Appendix~\ref{sec:dissCoupling}). Moreover, dissipative coupling cannot explain the observed scaling of resonant RIN ($\propto (G/\kappa)^4$) and the absence of correlation between the RIN level and the excess optical loss added by the membrane.

\subsection{Thermal intermodulation noise caused by a soft-clamped phononic crystal membrane}\label{sec:softClamped}

Localized (``soft-clamped") defect modes in stressed phononic crystal (PnC) resonators can have quality factors in excess of $10^8$ at room temperature due to enhanced dissipation dilution \cite{tsaturyan_ultracoherent_2017,ghadimi_strain_2017}. Owing to their high $Q$ and low effective mass, which result in low thermal force noise, $S_\t{FF,th}= 2k_B T m_\t{eff}\Gamma_m$, these modes are promising for quantum optomechanics experiments \cite{rossi_measurement-based_2018}.

In \figref{fig:softClamped}a and b we present \SiN PnC membranes with soft-clamped modes optimized for low effective mass and high $Q$. The phononic crystals are formed by the hexagonal pattern of circular holes introduced in Ref.~\cite{tsaturyan_ultracoherent_2017}, which creates a bandgap for flexural modes. The phononic crystal is terminated to the silicon frame at half the hole radii in order to prevent mode localization at the membrane edges---such modes have low $Q$ and can have frequencies within the phononic bandgap, contaminating the spectrum. \figref{fig:softClamped}a shows a microscope image of a resonator with a trampoline defect, featuring $m_\t{eff}=3.8$ ng and $Q=1.65\times 10^8$ at $0.853$ MHz, corresponding to a thermal force noise $S_\t{FF,th}=13$ aN/$\sqrt{\text{Hz}}$. Another resonator, shown in \figref{fig:softClamped}b, is a \mbrsize{2} phononic crystal membrane with a defect engineered to create a single mode localized in the middle of the phononic bandgap. The displayed sample has $Q=7.4\times 10^7$ at $1.46$ MHz and $m_\t{eff}=1.1$ ng, corresponding to $S_\t{FF,th}=34$ aN/$\sqrt{\text{Hz}}$.

The phononic bandgap spectrally isolates soft-clamped modes from the thermomechanical noise created by the rest of the membrane spectrum. Nevertheless, when a PnC membrane is incorporated in a MIM cavity the entire multitude of membrane modes contributes to the TIN \emph{even within bandgap frequencies}, as TIN is produced by a nonlinear process.
\figref{fig:detuningSweep}a shows the spectrum of light transmitted through a resonance of membrane-in-the-middle cavity with $g_0/2\pi=0.9$ kHz for the soft-clamped mode, $\kappa/2\pi=34$ MHz and $C_0=2.5$. The noise at bandgap frequencies is dominated by TIN, which exceeds the shot noise by four orders of magnitude. The spectrum also shows a dispersive feature in the middle of the bandgap, which is a signature of classical correlations due to the intracavity TIN exciting the localized mechanical mode. The mechanical resonator in this case is a 2~mm square PnC membrane with the patterning shown in \figref{fig:softClamped}b, but made of 40 nm-thick \SiN. The membrane has a single soft-clamped mode with $Q=4.1\times 10^7$ at $1.55$ MHz, as characterized immediately before inserting the membrane in the cavity assembly. The input power in the measurement was 100 $\mathrm{\mu W}$ after correcting for spatial mode matching, which corresponds to a nominal $C_q\approx 1$. The shot noise level was calibrated in a separate measurement by directing an independent laser beam on the detector.

We next present in \figref{fig:detuningSweep}b the dependence of the bandgap noise level on the laser detuning, measured on a different optical resonance of the same MIM cavity and at lower input power. The measurement shows that the in-bandgap excess noise is dominated by TIN at all detunings except for the immediate vicinity of the ``magic" detuning $\nu_0=-1/\sqrt{3}$. Around $\nu_0=-1/\sqrt{3}$ the excess noise is consistent with the substrate noise of an empty cavity (see Appendix~\ref{sec:expDetails}). The total noise level is well fitted by our model that includes both $S_{\nu\nu}$ and $S_{\nu\nu}^{(2)}$ contributions to the detected signal and accounts for the radiation pressure cooling (full details are given in Appendix~\ref{sec:detDep}). In the measurement in \figref{fig:detuningSweep}b $g_0/2\pi=360$ Hz for the localized mode, $\kappa/2\pi=24.8$ MHz and the input power was 30 $\mu$W.

Notice that the intensity of the detected light in our measurement is proportional to the intensity of the intracavity field. Therefore, the suppresion of TIN at magic detuning necessarily implies the suppression of the corresponding radiation pressure noise, which can lead to classical heating of the mechanical oscillator and thereby limit the true quantum cooperativity.

\section{A scheme to cancel intermodulation noise in detection}
The existence of a ``magic" detuning at which TIN is canceled in direct detection hints at a more general scheme. We show next that TIN can be removed from the measurement of an arbitrary optical quadrature at arbitrary cavity detuning by an appropriate choice of local oscillator in single-port homodyne detection. Single-port homodyning is a quantum-limited detection scheme if the signal beam is combined with local oscillator on a highly asymmetric beamsplitter.

Keeping terms up to the second order in $\delta \nu$, the complex amplitude of the optical field, $s$, in front of the photodetector of single-port homodyne is given by
\begin{equation}
s=s_0+s_1\,\delta \nu+s_2\,\delta \nu^2,
\end{equation}
where $s_1\propto L'(\nu_0)$, $s_2\propto L''(\nu_0)$. The carrier, $s_0=r e^{i\theta}$, can be set arbitrarily by an appropriate choice of the local oscillator phase and amplitude. The photocurrent is found as
\begin{equation}
I=|s_0|^2+(s_0^*s_1+s_0s_1^*)\delta \nu+(s_0^*s_2+s_0s_2^*+|s_1|^2)\delta \nu^2.
\end{equation}
At a given detection quadrature $\theta$, it is always possible to null the coefficient in front of the quadratic term, $\delta \nu^2$, by choosing the local oscillator such that
\begin{equation}
r=-|s_1|^2/(e^{-i\theta} s_2+ e^{i\theta} s_2^*).
\end{equation}
In this way the quadratic signal is absent from the measurement record and the nonlinearity of cavity transduction is in effect canceled by the nonlinearity of photodetection.

\section{Relation to nonlinear quantum measurements}
When a cavity is used as a meter to perform measurements on quantum systems \cite{clerk_introduction_2010}, the intracavity photon number, $\hat{n}_c$, is coupled to an observable of the system, $\hat{x}$, by means of the interaction Hamiltonian
\begin{equation}\label{eq:Hint}
\hat{H}_\t{int}=\hbar \omega_c(\hat{x})\,\hat{n}_c,
\end{equation}
where $\omega_c$ is the cavity frequency. The detected variable of the meter is the propagating optical field.
If coupling between the system and the meter is weak, the only way of performing nonlinear measurements, i.e. the measurement of $\hat{x}^2$, is creating non-linear coupling $\omega_{c}\propto x^2$ \cite{martin_measurement_2007,gangat_phonon_2011}.
In quantum optomechanics, where the microscopic system is a harmonic oscillator and $x$ is its position, quadratic measurements allow quantum non-demolition measurements of the oscillator energy \cite{braginsky_quantum_1992}. Such measurements can be used for the observation of phononic jumps \cite{martin_measurement_2007,gangat_phonon_2011}, phononic shot noise \cite{clerk_quantum_2010}, and the creation of mechanical squeezed states \cite{nunnenkamp_cooling_and_squeezing_2010} if the effects of linear measurement backaction are kept small \cite{martin_measurement_2007,brawley_nonlinear_2016}.
While considerable efforts have been dedicated to realizing nonlinear optomechanical coupling, achievable coupling rates remain modest \cite{thompson_strong_2008,paraiso_position-squared_2015}, and the corresponding experiments so far have been deeply in the classical regime.

A different path towards non-linear measurements \cite{vanner_selective_2011} opens as the sensitivity of meter is increased, for example by increasing the finesse of optical cavity.
Eventually measurements enter the regime when the information is still obtained gradually, but the first order perturbation theory \cite{clerk_quantum-limited_2004} is not sufficient to describe the measurement process \cite{lemonde_nonlinear_2013}.
In this regime, nonlinear measurements are possible with linear coupling, $\omega_{c}\propto x$, which was experimentally demonstrated \cite{brawley_nonlinear_2016,leijssen_nonlinear_2017}, yet in the classical domain. At the same time it was shown in \cite{brawley_nonlinear_2016}, and as we also show in this work (see Appendix~\ref{sec:quadOptom}), that under quite typical experimental conditions the nonlinearity of cavity as a meter is orders of magnitude stronger than the nonlinearity due to quadratic coupling.
Notably, it crucially requires the cavity being coupled to the propagating field, and does not have an equivalent in a closed system of a mechanical oscillator coupled to an isolated optical mode \cite{matsko_electromagnetic-continuum-induced_2018}.

\section{Conclusions and outlook}
We have presented the observation and characterization of a previously unreported broadband thermal noise, TIN, which originates in optical cavities from the quadratic transduction of frequency noise. Although produced by cavity frequency fluctuations, TIN is not correlated with them (neglecting radiation pressure effects) and therefore in many ways behaves as an independent noise. The key qualitative feature of TIN is that it creates intensity fluctuations in an optical field resonant with the cavity. The TIN magnitude grows quadratically with the ratio of r.m.s. thermal frequency fluctuations by the optical linewidth, and therefore it strongly affects high-finesse optical cavities with large frequency fluctuations, such as micro-cavities at room temperature.

Thermal intermodulation noise in optomechanical experiments can be avoided by using cavities with low finesse (equivalently, low $g_0/\kappa$), and by coupling them to mechanical resonators with lower total thermal fluctuations, i.e. which have fewer mechanical modes, higher frequency, and higher $Q$ for all modes. The latter consideration could make the fundamental modes of mechanical resonators (e.g. low-mass trampolines \cite{reinhardt_ultralow-noise_2016}) seem preferable compared to high-$Q$ but high-order PnC defect soft-clamped modes. In this context, a newly proposed method of exploiting self-similar structures as mechanical resonators with soft-clamped fundamental modes \cite{fedorov_fractal-like_2020} could potentially be fruitful for overcoming TIN. Another way of reducing the TIN is laser cooling of mechanical motion, either by dynamical backaction of a red-detuned beam or by active feedback. In this case, however, all mechanical modes that contribute to the total cavity noise must be efficiently cooled, which could be technically challenging.

The raw measurement data, analysis scripts and membrane designs are available in \cite{zenodo_repos}.

\section{Acknowledgements}
The authors thank Ryan Schilling for fabrication advice, Ramin Lalezari at FiveNine Optics for his assistance with the development of the cavity mirrors and Itay Shomroni for a careful reading of the manuscript. All samples were fabricated and grown in the Center of MicroNanoTechnology (CMi) at EPFL. This work was supported by the Swiss National Science Foundation under grant no. 182103 and the Defense Advanced Research Projects Agency (DARPA), Defense Sciences Office (DSO), under contract no. D19AP00016 (QUORT). A.B. acknowledges support from the European Union's Horizon 2020 research and innovation program under the Marie Sklodowska-Curie grant agreement no. 722923 (OMT). N.J.E. acknowledges support from the Swiss National Science Foundation under grant no. 185870 (Ambizione).


\appendix

\section{Experimental details}
\label{sec:expDetails}

\begin{figure*}[t]
\includegraphics[width=\textwidth]{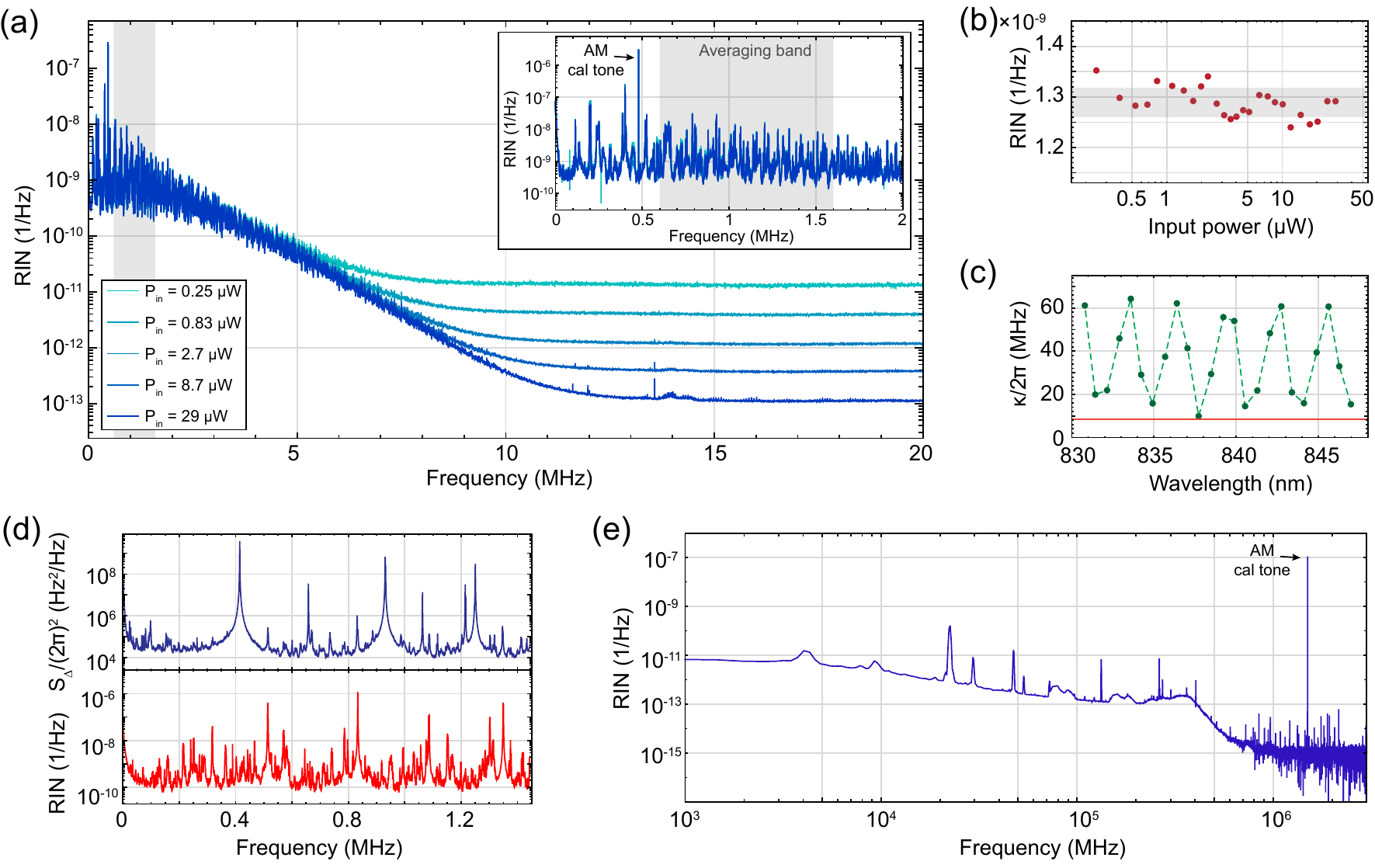}
\centering
\caption{Thermal noises in direct detection, classical laser noise and the variation of optical quality factors with wavelength. \small{a) Spectra of resonant relative intensity noise for a 2 mm square unpatterned membrane (resonance wavelength 837.7 nm, $g_0/2\pi=84$ Hz, $\kappa/2\pi=9.9$ MHz) at different input powers. The inset shows the same plot zoomed in at low frequencies. The RIN levels plotted in \figref{fig:rectMbr}b of main manuscript are averaged over the frequency range shaded gray. b) A reproduction of the average RIN from \figref{fig:rectMbr}b of main text. c)  Green points---measured linewidths of different optical resonances of MIM cavity with a 2mm$\times$2mm$\times$20nm unpatterned membrane, the dashed line is a guide to eye. Orange line---linewidth of an empty cavity with the same length. d) Low frequency zoom-in of the data in \figref{fig:rectMbr} of the main text. e) Relative intensity noise of the Ti:Sa laser used in the experiments, measured at 1 mW power.}}
\label{fig:extRectMbr}
\end{figure*}

Our experimental setup, as shown in \figref{fig:intro}c, comprises a membrane-in-the-middle cavity, consisting of two high-reflectivity, dielectric mirrors with 100 ppm transmission and a 200 $\mu$m-thick silicon chip which is sandwiched directly between the mirrors and hosts a suspended high-stress stoichiometric \SiN membrane. The total length of the cavity is around 350 $\mu$m, the cavity beam waist for the TEM$_{00}$ mode is about $35$ $\mu$m. The MIM cavity is situated in a vacuum chamber at room temperature and probed using a Ti:Sa or a tunable external cavity diode laser at a wavelength around 840 nm, close to the maximum reflectivity wavelength of the mirrors. A Ti:Sa laser was used in all the thermal noise measurements, whereas a diode laser was used for characterization of optical linewidths. The measurement signal was generated by direct detection of the light transmitted through the cavity on an avalanche photodiode. The reflected light, separated using a circulator, was used for Pound-Drever-Hall (PDH) locking of the Ti:Sa frequency. The one-sided spectra \cite{specta_notations} of signals were detected in transmission and calibrated either as relative intensity noise (RIN) or as effective cavity detuning fluctuations, $S_\Delta$, with the help of calibration tones applied to the amplitude or phase quadratures of the laser, respectively. Optomechanical vacuum coupling rates, $g_0$, were measured using frequency noise calibration as described in Ref.~\cite{gorodetksy_determination_2010}.

The characterization of TIN in \secref{sec:rectMbr} was performed using 20 nm-thick, square membranes with different side lengths as mechanical resonators. The insertion of a membrane into the cavity resulted in excess loss for most of the optical resonances. Nevertheless, for some resonances, the optical quality factors were reduced by only 10\% (a typical variation of optical loss with wavelength is shown in \figref{fig:extRectMbr}c). The optomechanical cooperativity was kept low during the noise measurements to eliminate dynamical backaction of the light (damping or amplification of mechanical motion). For this purpose the residual pressure in the vacuum chamber was kept high, $0.22\pm 0.03$ mBar, such that the quality factors of the fundamental modes of the membranes were limited by gas damping to $Q\sim 10^3$. The negligibility of dynamical backaction is verified by the data in \figref{fig:extRectMbr}a and b, which shows that the resonant RIN is constant when the input power is changed by two orders of magnitude. The spectra in \figref{fig:extRectMbr}a, taken at different powers, also show that at high frequencies (greater than 15 MHz), the shot noise is dominant within the measurement power range, which justifies the estimate of the shot noise levels in \figref{fig:rectMbr}d and e.

\begin{figure*}[t]
\includegraphics[width=\textwidth]{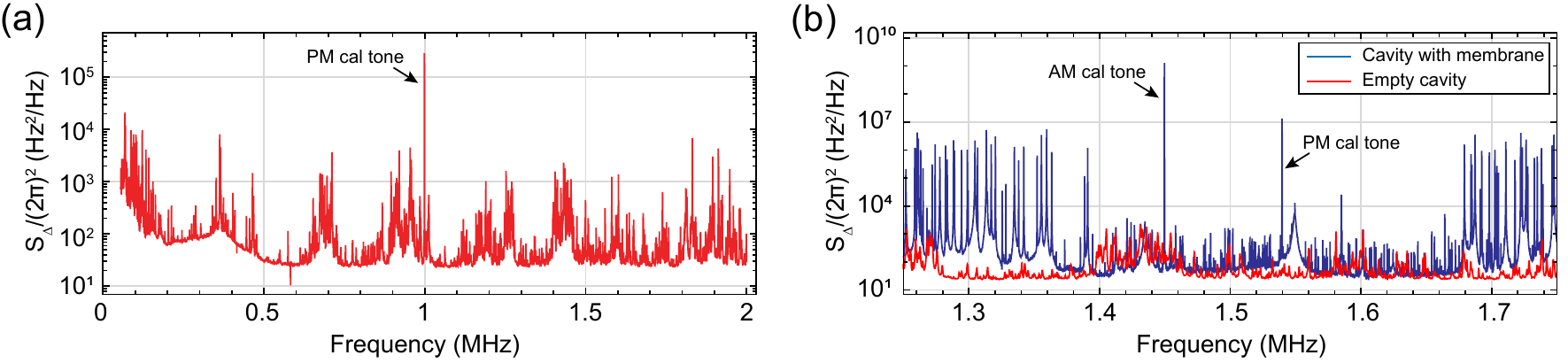}
\centering
\caption{\small{a) Spectrum of thermal detuning fluctuations due to the substrate noise, measured for a Fabry-Perot cavity without membrane. b) Substrate noise overlapped with a trace from detuning sweep presented in \figref{fig:detuningSweep}b of the main text corresponding to $2\Delta/\kappa=-0.51$.}}
\label{fig:mirrorNoise}
\end{figure*}

In all measurements, the classical intensity noise of the Ti:Sa laser (shown in \figref{fig:extRectMbr}e) was more than an order of magnitude below the resonant RIN of MIM cavities for frequencies above $100$ kHz. The classical frequency noise of the laser is below 1 Hz$^2$/Hz for frequencies above $100$ kHz (see the supplementary information of \cite{sudhir_quantum_2017}), which is at least an order of magnitude below the thermal detuning noise of an empty Fabry-Perot cavity (assembled using a silicon chip without a membrane as a spacer) as shown in \figref{fig:mirrorNoise}a. Additionally, we did not observe any significant effect of the laser lock performance on the magnitude of TIN, which indicates that the up-conversion of detuning noise from low frequencies (below $100$ kHz), where the laser noise is largest, contributes negligibly to the TIN in our cavities.

The measurements with PnC membrane presented in \secref{sec:softClamped} were made using the same setup shown in \figref{fig:intro}c, but in this case the vacuum pressure was kept below $5\times 10^{-7}$ mBar in order to eliminate gas damping. The insertion of a PnC membrane in a MIM cavity typically resulted in a somewhat larger excess optical loss than the insertion of a rectangular membrane. The excess optical loss was estimated to be 300 ppm per roundtrip for the cavity resonance in \figref{fig:detuningSweep}a, and 150 ppm for the resonance in \figref{fig:detuningSweep}b. For taking the data in \figref{fig:detuningSweep}b, the detuning of the laser from the cavity resonance was controlled by and inferred from the locking offset. For detunings greater than $2\Delta/\kappa\approx 0.5$, where the PDH error flips sign, side of the line locking was used instead of PDH. The bandgap noise was averaged over a 35 kHz band indicated in \figref{fig:detuningSweep}a.

\figref{fig:mirrorNoise}b shows an overlap of the detuning fluctuations spectra taken for an empty cavity and a cavity with PnC membrane at magic detuning, which shows an almost perfect coincidence of the extraneous noise levels. The MIM data corresponds to one of the minimum noise points in \figref{fig:detuningSweep}b of the main text and the membrane parameters can be found in the corresponding description. The uncertainty of the empty cavity noise indicated in \figref{fig:detuningSweep}b by a band comes from the fine structure of the noise peaks, which makes the exact level dependent on the selection of integration band.

\section{Membrane fabrication}

Patterned and unpatterned membrane samples are fabricated on the same \SI{100}{\milli \meter} wafer. Stoichiometric, high stress \ch{Si_3N_4} is grown by low pressure chemical vapor deposition (LPCVD) on both sides of a \SI{200}{\micro \meter}-thick silicon wafer. The initial deposition stress is estimated a posteriori from the observation of membrane resonant frequencies, and varies in the range \SI{900}{}-\SI{1100}{MPa}, changing slightly with deposition run.

The fabrication process relies on bulk wet etching of silicon in \ch{KOH} through the whole wafer thickness, to create openings for optical access to the membrane samples \cite{tsaturyan_ultracoherent_2017,reinhardt_ultralow-noise_2016,gartner2018integrated}. The extremely high selectivity of \ch{Si_3N_4} to \ch{Si} during \ch{KOH} etching allows the use of the backside nitride layer as a mask, to define the outline of the membranes on the frontside.

\begin{figure}[t]
\includegraphics[width=\columnwidth]{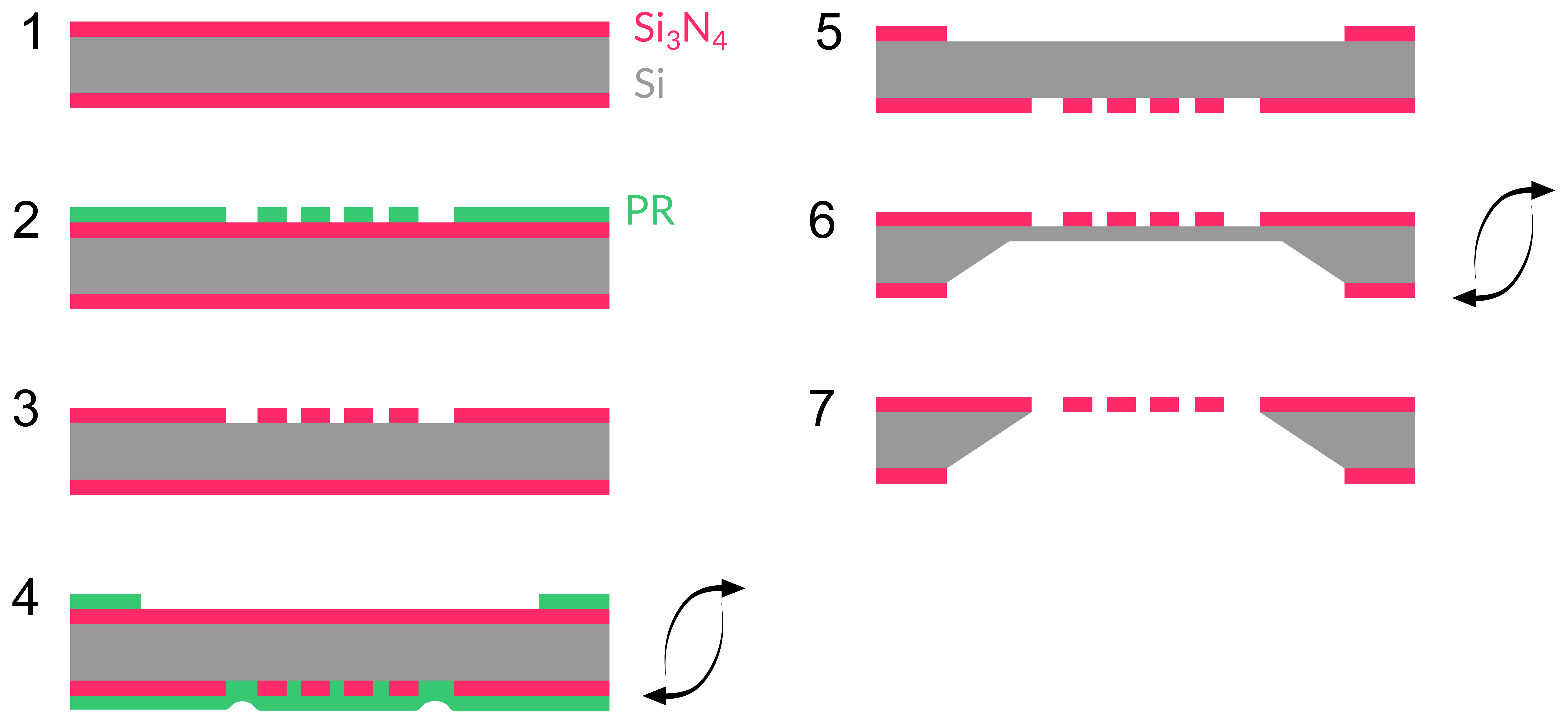}
\centering
\caption{\small{Main steps of the fabrication process. Magenta - \ch{Si_3N_4}; gray - Si; green - photoresist.}}
\label{fig:processflow}
\end{figure}

Initially, the frontside nitride (\ch{Si_3N_4}) layer is patterned with h-line photolithography and \ch{CHF_3}/\ch{SF6}-based reactive ion etching (RIE) (steps 2-3 of figure \ref{fig:processflow}). The photoresist film is then stripped with a sequence of hot N-Methyl-2-pyrrolidone (NMP) and \ch{O_2} plasma; this procedure is carefully repeated after each etching step. The frontside nitride layer is then protected by spinning a thick layer of negative-tone photoresist (MicroChemicals AZ\textregistered 15nXT), prior to flipping the wafer and beginning the patterning of membrane windows on the backside nitride layer (steps 4-5). We noticed a reduction in the occurrence of local defects and increased overall membrane yield when the unreleased membranes on the frontside were protected from contact with hot plates, spin-coaters and plasma etcher chucks. The backside layer is then patterned with membrane windows, in a completely analogous way. The exposure step requires a wafer thickness-dependent rescaling of membrane windows, to account for the slope of slow-etching planes in \ch{KOH}, and careful alignment with frontside features.

After stripping the photoresist, the wafer is installed in a PTFE holder for the first wet etching step in \ch{KOH} at $\approx \SI{75}{\celsius}$ (step 6). The holder clamps the wafer along its rim, sealing off the wafer frontside with a rubber O-ring, while exposing the backside to chemical etching by \ch{KOH}. This procedure is necessary to ensure that PnC membranes are suspended correctly: we noticed that releasing PnC samples by etching from both sides of the wafer produced a large number of defects in the phononic crystal, probably due to the particular dynamics of undercut and stress relaxation in the film. The wafer is etched until \SI{30}{}-\SI{40}{\micro\meter} of silicon remain, ensuring a sufficient sturdiness of the samples during the subsequent fabrication steps. The wafer is then removed from the KOH bath and the PTFE holder, rinsed and cleaned in concentrated \ch{HCl} at room temperature for 2 hours \cite{nielsen2004particle}.

Subsequently, the wafer is coated with thick, protective photoresist and diced into $\SI{8.75}{mm}\times\SI{8.75}{mm}$ chips, and the remainder of the process is carried on chip-wise. The chips are cleaned again with hot solvents and \ch{O_2} plasma, and the membrane release is completed by exposing the chips to \ch{KOH} from both sides (step 7). The temperature of the solution is lowered ($\approx \SI{55}{}-\SI{60}{\celsius}$), to mitigate the perturbation of fragile samples by buoyant \ch{N_2} bubbles, byproducts of the etching reaction. After the undercut is complete, the samples are carefully rinsed, cleaned in \ch{HCl}, transferred to an ethanol bath and gently dried in a critical point dryer (CPD).

\section{Quadratic mechanical displacement transduction by the optical cavity versus quadratic optomechanical coupling}\label{sec:quadOptom}

Nonlinear cavity transduction can produce signals quadratic in mechanic displacement which are orders of magnitude stronger than previously experimentally demonstrated quadratic coupling arising from $\partial^2 \omega_c/\partial x^2$ terms \cite{brawley_nonlinear_2016}. Below we derive the classical dynamics of the optical field in an optomechanical cavity taking into account terms that are quadratic in displacement. We show that in a membrane-in-the-middle cavity, the quadratic signals originating from nonlinear transduction are $r\mathcal{F}$ larger than the signals due to the nonlinear optomechanical coupling, $\partial^2 \omega_c/\partial x^2$.

The fluctuations of $\nu$ due to the mechanical displacement are given by
\begin{equation}
\delta\nu(t)\approx 2\frac{G}{\kappa} x(t)+\frac{G_2}{\kappa} x(t)^2,
\end{equation}
where $G=-\partial \omega_c/\partial x$ and $G_2=-\partial^2 \omega_c/\partial^2 x$ are the linear and quadratic optomechanical coupling, respectively, and the total displacement $x$ consists of partial contributions of different modes $x_n$
\begin{equation}
x(t) = \sum_n x_n(t).
\end{equation}
For a resonant laser probe we can find the intracavity field as
\begin{multline}\label{eq:aResDisp}
a(t)\approx 2\sqrt{\frac{\eta_1}{\kappa}}(1 -i\nu(t) -\nu(t)^2) s_\t{in,1}=\\
2\sqrt{\frac{\eta_1}{\kappa}}\left(1 -2i\frac{G}{\kappa} x(t) -\left(\left(2\frac{G}{\kappa} \right)^2+i\frac{G_2}{\kappa}\right) x(t)^2\right) s_\t{in,1}.
\end{multline}
It is instructive to compare the magnitudes of the two contributions to the prefactor of $x(t)^2$. The typical value for $G$ (assuming the membrane is not very close to one of the mirrors) is
\begin{equation}
G\sim 2r \frac{\omega_c}{l_c},
\end{equation}
while the typical value for $G_2$ is \cite{thompson_strong_2008}
\begin{equation}
G_2\sim 4 \frac{r \omega_c^2}{c \, l_c},
\end{equation}
where $c$ is the speed of light, $r$ is the membrane reflectivity and $l_c$ is the cavity length. The ratio of the two contributions is evaluated as
\begin{equation}
\left. \left(2\frac{G}{\kappa} \right)^2\right/\left(\frac{G_2}{\kappa}\right)\sim\mathcal{F}r.
\end{equation}
As the cavity finesse $\mathcal{F}$ is typically large, on on the order of $10^3$ to $10^5$, and the membrane reflectivity $r$ is between $0.1$ and $0.5$, we conclude that linear optomechanical coupling needs to extremely well suppressed in order for the quadratic coupling $G_2$ to contribute.

\section{Dissipative coupling}\label{sec:dissCoupling}
In an optomechanical membrane-in-the-middle cavity dissipative coupling, $\partial \kappa/\partial x$, exists in addition to the dispersive coupling, $\partial \omega_c/\partial x$. Dissipative coupling modulates the optical decay rate, both external coupling and intrinsic loss, and can potentially produce intensity noise in a resonantly locked probe laser. However, for the parameters of our experiment the dissipative coupling is negligible.

The noise due to dissipative coupling can be upper-bound as follows. The cavity linewidth cannot change by more than $\kappa$ as the membrane is translated by $\lambda$ inside the cavity, and therefore the dissipative coupling rate is limited by
\begin{equation}
\frac{\partial \kappa}{\partial x}\lesssim \frac{\kappa}{\lambda}= \frac{1}{\mathcal{F}}\frac{\omega_c}{2l_c}\sim \frac{G}{\mathcal{F}},
\end{equation}
where in the last transition it was assumed that the membrane reflectivity is of order unity.

The resonant intracavity field modulated by dissipative coupling is given by
\begin{equation}
a(t)\approx 2\sqrt{\frac{\eta_1}{\kappa}}\left(1 - \frac{1}{2\kappa}\frac{\partial \kappa}{\partial x} x(t)\right) s_\t{in,1}.
\end{equation}
Comparing to \eqref{eq:aResDisp}, we find that the noise produced by dissipative coupling is negligible compared to the intermodulation noise if
\begin{equation}
\frac{G}{\kappa}x \gg \frac{1}{\mathcal{F}}.
\end{equation}
In all the experiments presented in this work this condition is satisfied, $Gx/\kappa$ ranges from $10^{-3}$ to $10^{-1}$, whereas $1/\mathcal{F}$ is always less than $10^{-4}$.

\section{Details of TIN modeling}\label{sec:theoryDetails}

\noindent For an optomechanical system, the cavity resonance frequency shifts due to a displacement field $\vec{u}(\vec r)$ is, to the first order in $\vec{u}(\vec r)$, given by \cite{gorodetksy_determination_2010}

\begin{equation}
\frac{\Delta\omega_c}{\omega_c} = \frac{1}{2}\frac{\int |\vec E{(\vec r)}|^2\nabla\epsilon(\vec r)\cdot \vec{u}(\vec r)}{\int |\vec E{(\vec r)}|^2\epsilon(\vec r)}
\end{equation}
 where $\epsilon(\vec r)$ is the dielectric constant. For membrane in the middle systems, the gradient of the dielectric constant is nonzero only on the boundaries of the membrane slab. Therefore, the resonance frequency shift and correspondingly the linear optomechanical coupling constant ($G$) is simply  proportional to the average of the displacement field over the cavity mode shape at the position of the membrane. Hence, the optomechanical coupling constant for different flexural modes of a membrane can be written in the form $G_n = \eta_n\cdot G$, where $n$ is an index running over the mechanical modes, $G$ is a constant and $\eta_n$ are the overlap factors, proportional to the average of the mode shape on the cavity mode. For a membrane (in the x-y plane) with flexural modes $\{u_n(x,y)\}$, by choosing $G$ to be equal to the coupling constant for the fundamental mode ($G_1$) the overlap factors have the form

 \begin{equation}
     \eta_n = \frac{\int u_n(x,y)I(x,y)\,dxdy}{\int u_1(x,y)I(x,y)\,dxdy},
 \end{equation}
where $u_1$ denotes the fundamental mode and $I(x,y) = |\vec E{(x,y)}|^2$ is the cavity mode shape. For the TEM$_{00}$ mode of a Fabry-Perot cavity the transverse mode profile is given by
\begin{equation}
    I(x,y) = \sqrt{\frac{2}{\pi w^2}} e^{-2((x-x_0)^2+(y-y_0)^2)/w^2}
\end{equation}
where $w$ is the waist of the mode at the position of the membrane and $x_0$ and $y_0$ are the position coordinates of the beam center on the membrane. We normalize the coupling constants to the value for the fundamental mode since in the actual experiment we calibrate the coupling constant using the fundamental mode. With $x_n(t)$ the amplitude of the $n$-th mode, the total fluctuations of the cavity normalized detuning are given by
\begin{equation}
\delta\nu(t)  = \frac{2G}{\kappa} \sum_n \eta_n x_n(t).
\end{equation}
Comparing to \eqref{eq:dnuOpt} of the main text, we can identify the total displacement as
\begin{equation}
x(t) = \sum_n \eta_n x_n(t).
\end{equation}
The spectrum of the linear fluctuations of $x$, $S_{xx}[\omega]$, is a linear combination of thermal spectra of each mode, $S_{xx,n}[\omega]$, due to the fact that the Brownian motion of different modes are statistically independent. 
The linear position fluctuations then are written as

\begin{equation}\label{eq:S_xx}
S_{xx}[\omega] = \sum_n \eta_n^2 S_{xx,n}[\omega].
\end{equation}
$S_{xx,n}[\omega]$ is also given by the fluctuation-dissipation theorem.

\begin{equation}
    S_{xx,n}[\omega] = \frac{2kT}{\omega}\Im{\{\chi_n[\omega]\}}
\end{equation}
where $\chi_n[\omega]$ is the susceptibility of mode $n$. The quadratic fluctuations of $x$ can be calculated using a relation similar to \eqref{eq:Snu2} of the main text

\begin{equation}\label{eq:S_xx2}
    S_{xx}^{(2)}[\omega] = \frac{1}{\pi}\int_{-\infty}^{\infty}S_{xx}[\omega']S_{xx}[\omega-\omega']d\omega',
\end{equation}
where we have dropped the delta function term which is irrelevant to the numerical model. Having computed the linear and quadratic spectra of displacement fluctuations, the linear and quadratic frequency fluctuations are also calculated accordingly using \eqref{eq:mbrFreqFluct} and \eqref{eq:intNoiseSxx2} of the main text, and finally the total photocurrent spectrum can be calculated from \eqref{eq:detectedNoise} (see \eqref{eq:App_F_S_II}).

For a thin high stress square membrane with the side length $L$, the flexural modes are approximately given by products of sine waves
\begin{equation}
    u_{nm} = \frac{2}{L}\sin{(\frac{2n\pi x}{L})}\sin{(\frac{2m\pi y}{L})},
\end{equation}
with the mode frequencies $\Omega_{nm} = \frac{\pi c}{L}\sqrt{n^2 + m^2}$, where  $c=\sqrt{\sigma/\rho}$ is the speed of acoustic waves in a film with density of $\rho$ and stress $\sigma$. The effective mass for all modes is equal to $M/4$, a quarter of the total mass of the membrane. In the data presented in \figref{fig:theor} of the main text, the membrane is surrounded by a low vacuum environment: damping is predominantly viscous, with a constant damping rate given by $\Gamma_{nm} = \Omega_{nm}/Q_{nm}$. Piecing it all together, the susceptibility of the mode $\left(n,m\right)$ is given by

\begin{equation}
    \chi_{nm}[\omega] = \frac{1}{M/4}\frac{1}{\Omega_{nm}^2 - \omega^2 - i\Gamma_{nm}\omega}.
\end{equation}
Finally, we can analytically calculate the linear spectrum using \eqref{eq:S_xx} and then find the TIN by numerically computing the convolution integral in \eqref{eq:S_xx2}.


\section{The model of detuning dependence of total output light noise for MIM cavity with PnC membrane}
\label{sec:detDep}
As shown in the main manuscript text, the intensity of light, $I(t)$, and therefore the photodiode signal, is related to the linear ($\delta\nu(t)$) and quadratic ($\delta\nu(t)^2$) fluctuations of the cavity frequency as
\begin{multline}
I(t)=|s_\t{out,2}(t)|^2\propto\\
|L(\nu_0)|^2 \left(1-\frac{2\nu_0}{1+\nu_0^2}\delta\nu(t)+\frac{3\nu_0^2-1}{(1+\nu_0^2)^2}\delta\nu(t)^2\right),
\end{multline}
where $\nu_0=2\Delta_0/\kappa$ is normalized detuning. The spectrum of intensity fluctuations of the output light is given by,
\begin{equation}\label{eq:App_F_S_II}
S_{II}[\omega]\propto \frac{4 \nu_0^2}{(1+\nu_0^2)^2} S_{\nu\nu}[\omega]+\frac{(3\nu_0^2-1)^2}{(1+\nu_0^2)^4} S^{(2)}_{\nu\nu}[\omega].
\end{equation}
In an optomechanical cavity operated at high input power $S_{\nu\nu}$ and $S^{(2)}_{\nu\nu}$ in general are detuning-dependent because of the laser cooling/amplification of mechanical motion.

In order to find the dependence of $S_{II}$ on $\Delta$ some specific assumptions need to be made about the operating regime and the frequency of interest. Considering the case of data in \figref{fig:detuningSweep}b of the main text, here the noise level is estimated at the bandgap frequency and therefore only the mirror noise is expected to contribute to $S_{\nu\nu}$. The mechanical modes of the mirrors are relatively weakly coupled to the intracavity light and therefore the dynamical backaction for them can be neglected, resulting in detuning-independent $S_{\nu\nu}$. The intermodulation noise contribution, on the contrary, is significantly affected by laser cooling. It is natural to suggest (and it is advocated for by the very good agreement of our conclusions with experimental data) that TIN at bandgap frequencies is dominated by the mixing products of resonant and off-resonant parts of the membrane thermomechanical spectrum. Dynamical backaction reduces the mechanical spectral density on resonance $\propto 1/\Gamma_\t{DBA}$, where $\Gamma_\t{DBA}$ is the optical damping rate and $\Gamma_\t{DBA}\gg \Gamma_m$ is assumed, and it does not affect the off-resonant spectral density. In the unresolved-sideband regime, which is typically well fulfilled in our measurements, the optical damping rate is given by
\begin{equation}
\Gamma_\t{DBA}= -32\frac{\Omega_m}{\kappa}\left(\frac{2g_0}{\kappa}\right)^2\frac{\nu_0}{(1+\nu_0^2)^3} \eta_1|s_\t{in,1}|^2,
\end{equation}
and under our assumptions the spectral density of quadratic frequency fluctuations at PnC bandgap frequencies follows the detuning dependence of $1/\Gamma_\t{DBA}$,
\begin{equation}
 S^{(2)}_{\nu\nu}\propto \frac{(1+\nu_0^2)^3}{|\nu_0|},
\end{equation}
for $\nu_0<0$.

Motivated by this consideration, the experimental data in \figref{fig:detuningSweep}b is fitted with the model
\begin{equation}
S_{II}\propto \frac{4 \nu_0^2}{(1+\nu_0^2)^2} C_1+\frac{1}{|\nu_0|}\frac{(3\nu_0^2-1)^2}{1+\nu_0^2} C_2,
\end{equation}
where $C_1$ and $C_2$ are free parameters. It was found that the model very well reproduces the observed variation of output noise with detuning and the value of $C_1$ found from the fit is indeed consistent with independently measured mirror noise, as shown in \figref{fig:mirrorNoise}.

\bibliography{references}

\begin{thebibliography}{57}%
\makeatletter
\providecommand \@ifxundefined [1]{%
 \@ifx{#1\undefined}
}%
\providecommand \@ifnum [1]{%
 \ifnum #1\expandafter \@firstoftwo
 \else \expandafter \@secondoftwo
 \fi
}%
\providecommand \@ifx [1]{%
 \ifx #1\expandafter \@firstoftwo
 \else \expandafter \@secondoftwo
 \fi
}%
\providecommand \natexlab [1]{#1}%
\providecommand \enquote  [1]{``#1''}%
\providecommand \bibnamefont  [1]{#1}%
\providecommand \bibfnamefont [1]{#1}%
\providecommand \citenamefont [1]{#1}%
\providecommand \href@noop [0]{\@secondoftwo}%
\providecommand \href [0]{\begingroup \@sanitize@url \@href}%
\providecommand \@href[1]{\@@startlink{#1}\@@href}%
\providecommand \@@href[1]{\endgroup#1\@@endlink}%
\providecommand \@sanitize@url [0]{\catcode `\\12\catcode `\$12\catcode
  `\&12\catcode `\#12\catcode `\^12\catcode `\_12\catcode `\%12\relax}%
\providecommand \@@startlink[1]{}%
\providecommand \@@endlink[0]{}%
\providecommand \url  [0]{\begingroup\@sanitize@url \@url }%
\providecommand \@url [1]{\endgroup\@href {#1}{\urlprefix }}%
\providecommand \urlprefix  [0]{URL }%
\providecommand \Eprint [0]{\href }%
\providecommand \doibase [0]{http://dx.doi.org/}%
\providecommand \selectlanguage [0]{\@gobble}%
\providecommand \bibinfo  [0]{\@secondoftwo}%
\providecommand \bibfield  [0]{\@secondoftwo}%
\providecommand \translation [1]{[#1]}%
\providecommand \BibitemOpen [0]{}%
\providecommand \bibitemStop [0]{}%
\providecommand \bibitemNoStop [0]{.\EOS\space}%
\providecommand \EOS [0]{\spacefactor3000\relax}%
\providecommand \BibitemShut  [1]{\csname bibitem#1\endcsname}%
\let\auto@bib@innerbib\@empty
\bibitem [{\citenamefont {{LIGO Scientific Collaboration and Virgo
  Collaboration}}(2016)}]{ligo_collaboration_observation_gw_2016}%
  \BibitemOpen
  \bibfield  {author} {\bibinfo {author} {\bibnamefont {{LIGO Scientific
  Collaboration and Virgo Collaboration}}},\ }\bibfield  {title} {\enquote
  {\bibinfo {title} {Observation of {Gravitational} {Waves} from a {Binary}
  {Black} {Hole} {Merger}},}\ }\href {\doibase 10.1103/PhysRevLett.116.061102}
  {\bibfield  {journal} {\bibinfo  {journal} {Physical Review Letters}\
  }\textbf {\bibinfo {volume} {116}},\ \bibinfo {pages} {061102} (\bibinfo
  {year} {2016})}\BibitemShut {NoStop}%
\bibitem [{\citenamefont {Sterr}\ \emph {et~al.}(2009)\citenamefont {Sterr},
  \citenamefont {Legero}, \citenamefont {Kessler}, \citenamefont {Schnatz},
  \citenamefont {Grosche}, \citenamefont {Terra},\ and\ \citenamefont
  {Riehle}}]{sterr_ultrastable_2009}%
  \BibitemOpen
  \bibfield  {author} {\bibinfo {author} {\bibfnamefont {U.}~\bibnamefont
  {Sterr}}, \bibinfo {author} {\bibfnamefont {T.}~\bibnamefont {Legero}},
  \bibinfo {author} {\bibfnamefont {T.}~\bibnamefont {Kessler}}, \bibinfo
  {author} {\bibfnamefont {H.}~\bibnamefont {Schnatz}}, \bibinfo {author}
  {\bibfnamefont {G.}~\bibnamefont {Grosche}}, \bibinfo {author} {\bibfnamefont
  {O.}~\bibnamefont {Terra}}, \ and\ \bibinfo {author} {\bibfnamefont
  {F.}~\bibnamefont {Riehle}},\ }\bibfield  {title} {\enquote {\bibinfo {title}
  {Ultrastable lasers: new developments and applications},}\ }in\ \href
  {\doibase 10.1117/12.825217} {\emph {\bibinfo {booktitle} {Time and
  {Frequency} {Metrology} {II}}}},\ Vol.\ \bibinfo {volume} {7431}\ (\bibinfo
  {publisher} {International Society for Optics and Photonics},\ \bibinfo
  {year} {2009})\ p.\ \bibinfo {pages} {74310A}\BibitemShut {NoStop}%
\bibitem [{\citenamefont {Ye}\ \emph {et~al.}(2008)\citenamefont {Ye},
  \citenamefont {Kimble},\ and\ \citenamefont {Katori}}]{ye_quantum_2008}%
  \BibitemOpen
  \bibfield  {author} {\bibinfo {author} {\bibfnamefont {Jun}\ \bibnamefont
  {Ye}}, \bibinfo {author} {\bibfnamefont {H.~J.}\ \bibnamefont {Kimble}}, \
  and\ \bibinfo {author} {\bibfnamefont {Hidetoshi}\ \bibnamefont {Katori}},\
  }\bibfield  {title} {\enquote {\bibinfo {title} {Quantum {State}
  {Engineering} and {Precision} {Metrology} {Using} {State}-{Insensitive}
  {Light} {Traps}},}\ }\href {\doibase 10.1126/science.1148259} {\bibfield
  {journal} {\bibinfo  {journal} {Science}\ }\textbf {\bibinfo {volume}
  {320}},\ \bibinfo {pages} {1734--1738} (\bibinfo {year} {2008})}\BibitemShut
  {NoStop}%
\bibitem [{\citenamefont {Aspelmeyer}\ \emph {et~al.}(2014)\citenamefont
  {Aspelmeyer}, \citenamefont {Kippenberg},\ and\ \citenamefont
  {Marquardt}}]{aspelmeyer_cavity_2014}%
  \BibitemOpen
  \bibfield  {author} {\bibinfo {author} {\bibfnamefont {Markus}\ \bibnamefont
  {Aspelmeyer}}, \bibinfo {author} {\bibfnamefont {Tobias~J.}\ \bibnamefont
  {Kippenberg}}, \ and\ \bibinfo {author} {\bibfnamefont {Florian}\
  \bibnamefont {Marquardt}},\ }\bibfield  {title} {\enquote {\bibinfo {title}
  {Cavity optomechanics},}\ }\href {\doibase 10.1103/RevModPhys.86.1391}
  {\bibfield  {journal} {\bibinfo  {journal} {Reviews of Modern Physics}\
  }\textbf {\bibinfo {volume} {86}},\ \bibinfo {pages} {1391} (\bibinfo {year}
  {2014})}\BibitemShut {NoStop}%
\bibitem [{\citenamefont {Clerk}\ \emph
  {et~al.}(2010{\natexlab{a}})\citenamefont {Clerk}, \citenamefont {Devoret},
  \citenamefont {Girvin}, \citenamefont {Marquardt},\ and\ \citenamefont
  {Schoelkopf}}]{clerk_introduction_2010}%
  \BibitemOpen
  \bibfield  {author} {\bibinfo {author} {\bibfnamefont {A.~A.}\ \bibnamefont
  {Clerk}}, \bibinfo {author} {\bibfnamefont {M.~H.}\ \bibnamefont {Devoret}},
  \bibinfo {author} {\bibfnamefont {S.~M.}\ \bibnamefont {Girvin}}, \bibinfo
  {author} {\bibfnamefont {Florian}\ \bibnamefont {Marquardt}}, \ and\ \bibinfo
  {author} {\bibfnamefont {R.~J.}\ \bibnamefont {Schoelkopf}},\ }\bibfield
  {title} {\enquote {\bibinfo {title} {Introduction to quantum noise,
  measurement, and amplification},}\ }\href {\doibase
  10.1103/RevModPhys.82.1155} {\bibfield  {journal} {\bibinfo  {journal}
  {Reviews of Modern Physics}\ }\textbf {\bibinfo {volume} {82}},\ \bibinfo
  {pages} {1155--1208} (\bibinfo {year} {2010}{\natexlab{a}})},\ \bibinfo
  {note} {publisher: American Physical Society}\BibitemShut {NoStop}%
\bibitem [{\citenamefont {Braginsky}\ \emph {et~al.}(2000)\citenamefont
  {Braginsky}, \citenamefont {Gorodetsky},\ and\ \citenamefont
  {Vyatchanin}}]{braginsky_thermorefractive_2000}%
  \BibitemOpen
  \bibfield  {author} {\bibinfo {author} {\bibfnamefont {V.~B.}\ \bibnamefont
  {Braginsky}}, \bibinfo {author} {\bibfnamefont {M.~L.}\ \bibnamefont
  {Gorodetsky}}, \ and\ \bibinfo {author} {\bibfnamefont {S.~P.}\ \bibnamefont
  {Vyatchanin}},\ }\bibfield  {title} {\enquote {\bibinfo {title}
  {Thermo-refractive noise in gravitational wave antennae},}\ }\href {\doibase
  10.1016/S0375-9601(00)00389-3} {\bibfield  {journal} {\bibinfo  {journal}
  {Physics Letters A}\ }\textbf {\bibinfo {volume} {271}},\ \bibinfo {pages}
  {303--307} (\bibinfo {year} {2000})}\BibitemShut {NoStop}%
\bibitem [{\citenamefont
  {Gorodetsky}(2008)}]{gorodetsky_thermal_noise_compensation_2008}%
  \BibitemOpen
  \bibfield  {author} {\bibinfo {author} {\bibfnamefont {Michael~L.}\
  \bibnamefont {Gorodetsky}},\ }\bibfield  {title} {\enquote {\bibinfo {title}
  {Thermal noises and noise compensation in high-reflection multilayer
  coating},}\ }\href {\doibase 10.1016/j.physleta.2008.09.056} {\bibfield
  {journal} {\bibinfo  {journal} {Physics Letters A}\ }\textbf {\bibinfo
  {volume} {372}},\ \bibinfo {pages} {6813--6822} (\bibinfo {year}
  {2008})}\BibitemShut {NoStop}%
\bibitem [{\citenamefont {Braginsky}\ \emph {et~al.}(1999)\citenamefont
  {Braginsky}, \citenamefont {Gorodetsky},\ and\ \citenamefont
  {Vyatchanin}}]{braginsky_thermodynamical_1999}%
  \BibitemOpen
  \bibfield  {author} {\bibinfo {author} {\bibfnamefont {V.~B.}\ \bibnamefont
  {Braginsky}}, \bibinfo {author} {\bibfnamefont {M.~L.}\ \bibnamefont
  {Gorodetsky}}, \ and\ \bibinfo {author} {\bibfnamefont {S.~P.}\ \bibnamefont
  {Vyatchanin}},\ }\bibfield  {title} {\enquote {\bibinfo {title}
  {Thermodynamical fluctuations and photo-thermal shot noise in gravitational
  wave antennae},}\ }\href {\doibase 10.1016/S0375-9601(99)00785-9} {\bibfield
  {journal} {\bibinfo  {journal} {Physics Letters A}\ }\textbf {\bibinfo
  {volume} {264}},\ \bibinfo {pages} {1--10} (\bibinfo {year}
  {1999})}\BibitemShut {NoStop}%
\bibitem [{\citenamefont {Cole}\ \emph {et~al.}(2013)\citenamefont {Cole},
  \citenamefont {Zhang}, \citenamefont {Martin}, \citenamefont {Ye},\ and\
  \citenamefont {Aspelmeyer}}]{cole_tenfold_2013}%
  \BibitemOpen
  \bibfield  {author} {\bibinfo {author} {\bibfnamefont {Garrett~D.}\
  \bibnamefont {Cole}}, \bibinfo {author} {\bibfnamefont {Wei}\ \bibnamefont
  {Zhang}}, \bibinfo {author} {\bibfnamefont {Michael~J.}\ \bibnamefont
  {Martin}}, \bibinfo {author} {\bibfnamefont {Jun}\ \bibnamefont {Ye}}, \ and\
  \bibinfo {author} {\bibfnamefont {Markus}\ \bibnamefont {Aspelmeyer}},\
  }\bibfield  {title} {\enquote {\bibinfo {title} {Tenfold reduction of
  brownian noise in high-reflectivity optical coatings},}\ }\href {\doibase
  10.1038/nphoton.2013.174} {\bibfield  {journal} {\bibinfo  {journal} {Nature
  Photonics}\ }\textbf {\bibinfo {volume} {7}},\ \bibinfo {pages} {644--650}
  (\bibinfo {year} {2013})}\BibitemShut {NoStop}%
\bibitem [{\citenamefont {Robinson}\ \emph {et~al.}(2019)\citenamefont
  {Robinson}, \citenamefont {Oelker}, \citenamefont {Milner}, \citenamefont
  {Zhang}, \citenamefont {Legero}, \citenamefont {Matei}, \citenamefont
  {Riehle}, \citenamefont {Sterr},\ and\ \citenamefont
  {Ye}}]{robinson_crystalline_2019}%
  \BibitemOpen
  \bibfield  {author} {\bibinfo {author} {\bibfnamefont {John~M.}\ \bibnamefont
  {Robinson}}, \bibinfo {author} {\bibfnamefont {Eric}\ \bibnamefont {Oelker}},
  \bibinfo {author} {\bibfnamefont {William~R.}\ \bibnamefont {Milner}},
  \bibinfo {author} {\bibfnamefont {Wei}\ \bibnamefont {Zhang}}, \bibinfo
  {author} {\bibfnamefont {Thomas}\ \bibnamefont {Legero}}, \bibinfo {author}
  {\bibfnamefont {Dan~G.}\ \bibnamefont {Matei}}, \bibinfo {author}
  {\bibfnamefont {Fritz}\ \bibnamefont {Riehle}}, \bibinfo {author}
  {\bibfnamefont {Uwe}\ \bibnamefont {Sterr}}, \ and\ \bibinfo {author}
  {\bibfnamefont {Jun}\ \bibnamefont {Ye}},\ }\bibfield  {title} {\enquote
  {\bibinfo {title} {Crystalline optical cavity at 4 k with
  thermal-noise-limited instability and ultralow drift},}\ }\href {\doibase
  10.1364/OPTICA.6.000240} {\bibfield  {journal} {\bibinfo  {journal} {Optica}\
  }\textbf {\bibinfo {volume} {6}},\ \bibinfo {pages} {240--243} (\bibinfo
  {year} {2019})},\ \bibinfo {note} {publisher: Optical Society of
  America}\BibitemShut {NoStop}%
\bibitem [{\citenamefont {Matsko}\ \emph {et~al.}(2007)\citenamefont {Matsko},
  \citenamefont {Savchenkov}, \citenamefont {Yu},\ and\ \citenamefont
  {Maleki}}]{matsko_whispering-gallery-mode_2007}%
  \BibitemOpen
  \bibfield  {author} {\bibinfo {author} {\bibfnamefont {Andrey~B.}\
  \bibnamefont {Matsko}}, \bibinfo {author} {\bibfnamefont {Anatoliy~A.}\
  \bibnamefont {Savchenkov}}, \bibinfo {author} {\bibfnamefont {Nan}\
  \bibnamefont {Yu}}, \ and\ \bibinfo {author} {\bibfnamefont {Lute}\
  \bibnamefont {Maleki}},\ }\bibfield  {title} {\enquote {\bibinfo {title}
  {Whispering-gallery-mode resonators as frequency references. i. fundamental
  limitations},}\ }\href {\doibase 10.1364/JOSAB.24.001324} {\bibfield
  {journal} {\bibinfo  {journal} {{JOSA} B}\ }\textbf {\bibinfo {volume}
  {24}},\ \bibinfo {pages} {1324--1335} (\bibinfo {year} {2007})},\ \bibinfo
  {note} {publisher: Optical Society of America}\BibitemShut {NoStop}%
\bibitem [{\citenamefont {Alnis}\ \emph {et~al.}(2011)\citenamefont {Alnis},
  \citenamefont {Schliesser}, \citenamefont {Wang}, \citenamefont {Hofer},
  \citenamefont {Kippenberg},\ and\ \citenamefont
  {Hänsch}}]{alnis_thermal-noise-limited_2011}%
  \BibitemOpen
  \bibfield  {author} {\bibinfo {author} {\bibfnamefont {J.}~\bibnamefont
  {Alnis}}, \bibinfo {author} {\bibfnamefont {A.}~\bibnamefont {Schliesser}},
  \bibinfo {author} {\bibfnamefont {C.~Y.}\ \bibnamefont {Wang}}, \bibinfo
  {author} {\bibfnamefont {J.}~\bibnamefont {Hofer}}, \bibinfo {author}
  {\bibfnamefont {T.~J.}\ \bibnamefont {Kippenberg}}, \ and\ \bibinfo {author}
  {\bibfnamefont {T.~W.}\ \bibnamefont {Hänsch}},\ }\bibfield  {title}
  {\enquote {\bibinfo {title} {Thermal-noise-limited crystalline
  whispering-gallery-mode resonator for laser stabilization},}\ }\href
  {\doibase 10.1103/PhysRevA.84.011804} {\bibfield  {journal} {\bibinfo
  {journal} {Physical Review A}\ }\textbf {\bibinfo {volume} {84}},\ \bibinfo
  {pages} {011804} (\bibinfo {year} {2011})},\ \bibinfo {note} {publisher:
  American Physical Society}\BibitemShut {NoStop}%
\bibitem [{\citenamefont {Lee}\ \emph {et~al.}(2013)\citenamefont {Lee},
  \citenamefont {Suh}, \citenamefont {Chen}, \citenamefont {Li}, \citenamefont
  {Diddams},\ and\ \citenamefont {Vahala}}]{lee_spiral_2013}%
  \BibitemOpen
  \bibfield  {author} {\bibinfo {author} {\bibfnamefont {Hansuek}\ \bibnamefont
  {Lee}}, \bibinfo {author} {\bibfnamefont {Myoung-Gyun}\ \bibnamefont {Suh}},
  \bibinfo {author} {\bibfnamefont {Tong}\ \bibnamefont {Chen}}, \bibinfo
  {author} {\bibfnamefont {Jiang}\ \bibnamefont {Li}}, \bibinfo {author}
  {\bibfnamefont {Scott~A.}\ \bibnamefont {Diddams}}, \ and\ \bibinfo {author}
  {\bibfnamefont {Kerry~J.}\ \bibnamefont {Vahala}},\ }\bibfield  {title}
  {\enquote {\bibinfo {title} {Spiral resonators for on-chip laser frequency
  stabilization},}\ }\href {\doibase 10.1038/ncomms3468} {\bibfield  {journal}
  {\bibinfo  {journal} {Nature Communications}\ }\textbf {\bibinfo {volume}
  {4}},\ \bibinfo {pages} {1--6} (\bibinfo {year} {2013})}\BibitemShut
  {NoStop}%
\bibitem [{\citenamefont {Dutt}\ \emph {et~al.}(2015)\citenamefont {Dutt},
  \citenamefont {Luke}, \citenamefont {Manipatruni}, \citenamefont {Gaeta},
  \citenamefont {Nussenzveig},\ and\ \citenamefont {Lipson}}]{dutt_-chip_2015}%
  \BibitemOpen
  \bibfield  {author} {\bibinfo {author} {\bibfnamefont {Avik}\ \bibnamefont
  {Dutt}}, \bibinfo {author} {\bibfnamefont {Kevin}\ \bibnamefont {Luke}},
  \bibinfo {author} {\bibfnamefont {Sasikanth}\ \bibnamefont {Manipatruni}},
  \bibinfo {author} {\bibfnamefont {Alexander~L.}\ \bibnamefont {Gaeta}},
  \bibinfo {author} {\bibfnamefont {Paulo}\ \bibnamefont {Nussenzveig}}, \ and\
  \bibinfo {author} {\bibfnamefont {Michal}\ \bibnamefont {Lipson}},\
  }\bibfield  {title} {\enquote {\bibinfo {title} {On-chip optical
  squeezing},}\ }\href {\doibase 10.1103/PhysRevApplied.3.044005} {\bibfield
  {journal} {\bibinfo  {journal} {Physical Review Applied}\ }\textbf {\bibinfo
  {volume} {3}},\ \bibinfo {pages} {044005} (\bibinfo {year} {2015})},\
  \bibinfo {note} {publisher: American Physical Society}\BibitemShut {NoStop}%
\bibitem [{\citenamefont {Hoff}\ \emph {et~al.}(2015)\citenamefont {Hoff},
  \citenamefont {Nielsen},\ and\ \citenamefont
  {Andersen}}]{hoff_integrated_2015}%
  \BibitemOpen
  \bibfield  {author} {\bibinfo {author} {\bibfnamefont {Ulrich~B.}\
  \bibnamefont {Hoff}}, \bibinfo {author} {\bibfnamefont {Bo~M.}\ \bibnamefont
  {Nielsen}}, \ and\ \bibinfo {author} {\bibfnamefont {Ulrik~L.}\ \bibnamefont
  {Andersen}},\ }\bibfield  {title} {\enquote {\bibinfo {title} {Integrated
  source of broadband quadrature squeezed light},}\ }\href {\doibase
  10.1364/OE.23.012013} {\bibfield  {journal} {\bibinfo  {journal} {Optics
  Express}\ }\textbf {\bibinfo {volume} {23}},\ \bibinfo {pages} {12013--12036}
  (\bibinfo {year} {2015})},\ \bibinfo {note} {publisher: Optical Society of
  America}\BibitemShut {NoStop}%
\bibitem [{\citenamefont {Otterpohl}\ \emph {et~al.}(2019)\citenamefont
  {Otterpohl}, \citenamefont {Otterpohl}, \citenamefont {Otterpohl},
  \citenamefont {Sedlmeir}, \citenamefont {Sedlmeir}, \citenamefont {Vogl},
  \citenamefont {Vogl}, \citenamefont {Dirmeier}, \citenamefont {Dirmeier},
  \citenamefont {Shafiee}, \citenamefont {Shafiee}, \citenamefont {Schunk},
  \citenamefont {Schunk}, \citenamefont {Schunk}, \citenamefont {Strekalov},
  \citenamefont {Schwefel}, \citenamefont {Schwefel}, \citenamefont {Gehring},
  \citenamefont {Andersen}, \citenamefont {Andersen}, \citenamefont {Leuchs},
  \citenamefont {Leuchs}, \citenamefont {Marquardt},\ and\ \citenamefont
  {Marquardt}}]{otterpohl_squeezed_2019}%
  \BibitemOpen
  \bibfield  {author} {\bibinfo {author} {\bibfnamefont {Alexander}\
  \bibnamefont {Otterpohl}}, \bibinfo {author} {\bibfnamefont {Alexander}\
  \bibnamefont {Otterpohl}}, \bibinfo {author} {\bibfnamefont {Alexander}\
  \bibnamefont {Otterpohl}}, \bibinfo {author} {\bibfnamefont {Florian}\
  \bibnamefont {Sedlmeir}}, \bibinfo {author} {\bibfnamefont {Florian}\
  \bibnamefont {Sedlmeir}}, \bibinfo {author} {\bibfnamefont {Ulrich}\
  \bibnamefont {Vogl}}, \bibinfo {author} {\bibfnamefont {Ulrich}\ \bibnamefont
  {Vogl}}, \bibinfo {author} {\bibfnamefont {Thomas}\ \bibnamefont {Dirmeier}},
  \bibinfo {author} {\bibfnamefont {Thomas}\ \bibnamefont {Dirmeier}}, \bibinfo
  {author} {\bibfnamefont {Golnoush}\ \bibnamefont {Shafiee}}, \bibinfo
  {author} {\bibfnamefont {Golnoush}\ \bibnamefont {Shafiee}}, \bibinfo
  {author} {\bibfnamefont {Gerhard}\ \bibnamefont {Schunk}}, \bibinfo {author}
  {\bibfnamefont {Gerhard}\ \bibnamefont {Schunk}}, \bibinfo {author}
  {\bibfnamefont {Gerhard}\ \bibnamefont {Schunk}}, \bibinfo {author}
  {\bibfnamefont {Dmitry~V.}\ \bibnamefont {Strekalov}}, \bibinfo {author}
  {\bibfnamefont {Harald G.~L.}\ \bibnamefont {Schwefel}}, \bibinfo {author}
  {\bibfnamefont {Harald G.~L.}\ \bibnamefont {Schwefel}}, \bibinfo {author}
  {\bibfnamefont {Tobias}\ \bibnamefont {Gehring}}, \bibinfo {author}
  {\bibfnamefont {Ulrik~L.}\ \bibnamefont {Andersen}}, \bibinfo {author}
  {\bibfnamefont {Ulrik~L.}\ \bibnamefont {Andersen}}, \bibinfo {author}
  {\bibfnamefont {Gerd}\ \bibnamefont {Leuchs}}, \bibinfo {author}
  {\bibfnamefont {Gerd}\ \bibnamefont {Leuchs}}, \bibinfo {author}
  {\bibfnamefont {Christoph}\ \bibnamefont {Marquardt}}, \ and\ \bibinfo
  {author} {\bibfnamefont {Christoph}\ \bibnamefont {Marquardt}},\ }\bibfield
  {title} {\enquote {\bibinfo {title} {Squeezed vacuum states from a whispering
  gallery mode resonator},}\ }\href {\doibase 10.1364/OPTICA.6.001375}
  {\bibfield  {journal} {\bibinfo  {journal} {Optica}\ }\textbf {\bibinfo
  {volume} {6}},\ \bibinfo {pages} {1375--1380} (\bibinfo {year} {2019})},\
  \bibinfo {note} {publisher: Optical Society of America}\BibitemShut {NoStop}%
\bibitem [{\citenamefont {Kippenberg}\ \emph {et~al.}(2018)\citenamefont
  {Kippenberg}, \citenamefont {Gaeta}, \citenamefont {Lipson},\ and\
  \citenamefont {Gorodetsky}}]{kippenberg_dissipative_2018}%
  \BibitemOpen
  \bibfield  {author} {\bibinfo {author} {\bibfnamefont {Tobias~J.}\
  \bibnamefont {Kippenberg}}, \bibinfo {author} {\bibfnamefont {Alexander~L.}\
  \bibnamefont {Gaeta}}, \bibinfo {author} {\bibfnamefont {Michal}\
  \bibnamefont {Lipson}}, \ and\ \bibinfo {author} {\bibfnamefont {Michael~L.}\
  \bibnamefont {Gorodetsky}},\ }\bibfield  {title} {\enquote {\bibinfo {title}
  {Dissipative {Kerr} solitons in optical microresonators},}\ }\href {\doibase
  10.1126/science.aan8083} {\bibfield  {journal} {\bibinfo  {journal}
  {Science}\ }\textbf {\bibinfo {volume} {361}} (\bibinfo {year} {2018}),\
  10.1126/science.aan8083}\BibitemShut {NoStop}%
\bibitem [{\citenamefont {Drake}\ \emph {et~al.}(2019)\citenamefont {Drake},
  \citenamefont {Stone}, \citenamefont {Briles},\ and\ \citenamefont
  {Papp}}]{drake_thermal_2019}%
  \BibitemOpen
  \bibfield  {author} {\bibinfo {author} {\bibfnamefont {Tara~E.}\ \bibnamefont
  {Drake}}, \bibinfo {author} {\bibfnamefont {Jordan~R.}\ \bibnamefont
  {Stone}}, \bibinfo {author} {\bibfnamefont {Travis~C.}\ \bibnamefont
  {Briles}}, \ and\ \bibinfo {author} {\bibfnamefont {Scott~B.}\ \bibnamefont
  {Papp}},\ }\bibfield  {title} {\enquote {\bibinfo {title} {Thermal
  decoherence and laser cooling of kerr microresonator solitons},}\ }\href
  {http://arxiv.org/abs/1903.00431} {\bibfield  {journal} {\bibinfo  {journal}
  {{arXiv}:1903.00431 [physics]}\ } (\bibinfo {year} {2019})},\ \Eprint
  {http://arxiv.org/abs/1903.00431} {1903.00431} \BibitemShut {NoStop}%
\bibitem [{\citenamefont {Gorodetsky}\ and\ \citenamefont
  {Grudinin}(2004)}]{gorodetsky_fundamental_2004}%
  \BibitemOpen
  \bibfield  {author} {\bibinfo {author} {\bibfnamefont {Michael~L.}\
  \bibnamefont {Gorodetsky}}\ and\ \bibinfo {author} {\bibfnamefont {Ivan~S.}\
  \bibnamefont {Grudinin}},\ }\bibfield  {title} {\enquote {\bibinfo {title}
  {Fundamental thermal fluctuations in microspheres},}\ }\href {\doibase
  10.1364/JOSAB.21.000697} {\bibfield  {journal} {\bibinfo  {journal} {{JOSA}
  B}\ }\textbf {\bibinfo {volume} {21}},\ \bibinfo {pages} {697--705} (\bibinfo
  {year} {2004})}\BibitemShut {NoStop}%
\bibitem [{\citenamefont {Anetsberger}\ \emph {et~al.}(2009)\citenamefont
  {Anetsberger}, \citenamefont {Arcizet}, \citenamefont {Unterreithmeier},
  \citenamefont {Rivière}, \citenamefont {Schliesser}, \citenamefont {Weig},
  \citenamefont {Kotthaus},\ and\ \citenamefont
  {Kippenberg}}]{anetsberger_near-field_2009}%
  \BibitemOpen
  \bibfield  {author} {\bibinfo {author} {\bibfnamefont {G.}~\bibnamefont
  {Anetsberger}}, \bibinfo {author} {\bibfnamefont {O.}~\bibnamefont
  {Arcizet}}, \bibinfo {author} {\bibfnamefont {Q.~P.}\ \bibnamefont
  {Unterreithmeier}}, \bibinfo {author} {\bibfnamefont {R.}~\bibnamefont
  {Rivière}}, \bibinfo {author} {\bibfnamefont {A.}~\bibnamefont
  {Schliesser}}, \bibinfo {author} {\bibfnamefont {E.~M.}\ \bibnamefont
  {Weig}}, \bibinfo {author} {\bibfnamefont {J.~P.}\ \bibnamefont {Kotthaus}},
  \ and\ \bibinfo {author} {\bibfnamefont {T.~J.}\ \bibnamefont {Kippenberg}},\
  }\bibfield  {title} {\enquote {\bibinfo {title} {Near-field cavity
  optomechanics with nanomechanical oscillators},}\ }\href {\doibase
  10.1038/nphys1425} {\bibfield  {journal} {\bibinfo  {journal} {Nature
  Physics}\ }\textbf {\bibinfo {volume} {5}},\ \bibinfo {pages} {909--914}
  (\bibinfo {year} {2009})}\BibitemShut {NoStop}%
\bibitem [{\citenamefont {Huang}\ \emph {et~al.}(2019)\citenamefont {Huang},
  \citenamefont {Lucas}, \citenamefont {Liu}, \citenamefont {Raja},
  \citenamefont {Lihachev}, \citenamefont {Gorodetsky}, \citenamefont
  {Engelsen},\ and\ \citenamefont {Kippenberg}}]{huang_thermorefractive_2019}%
  \BibitemOpen
  \bibfield  {author} {\bibinfo {author} {\bibfnamefont {Guanhao}\ \bibnamefont
  {Huang}}, \bibinfo {author} {\bibfnamefont {Erwan}\ \bibnamefont {Lucas}},
  \bibinfo {author} {\bibfnamefont {Junqiu}\ \bibnamefont {Liu}}, \bibinfo
  {author} {\bibfnamefont {Arslan~S.}\ \bibnamefont {Raja}}, \bibinfo {author}
  {\bibfnamefont {Grigory}\ \bibnamefont {Lihachev}}, \bibinfo {author}
  {\bibfnamefont {Michael~L.}\ \bibnamefont {Gorodetsky}}, \bibinfo {author}
  {\bibfnamefont {Nils~J.}\ \bibnamefont {Engelsen}}, \ and\ \bibinfo {author}
  {\bibfnamefont {Tobias~J.}\ \bibnamefont {Kippenberg}},\ }\bibfield  {title}
  {\enquote {\bibinfo {title} {Thermorefractive noise in silicon-nitride
  microresonators},}\ }\href {\doibase 10.1103/PhysRevA.99.061801} {\bibfield
  {journal} {\bibinfo  {journal} {Physical Review A}\ }\textbf {\bibinfo
  {volume} {99}},\ \bibinfo {pages} {061801} (\bibinfo {year} {2019})},\
  \bibinfo {note} {publisher: American Physical Society}\BibitemShut {NoStop}%
\bibitem [{\citenamefont {Vanner}(2011)}]{vanner_selective_2011}%
  \BibitemOpen
  \bibfield  {author} {\bibinfo {author} {\bibfnamefont {Michael~R.}\
  \bibnamefont {Vanner}},\ }\bibfield  {title} {\enquote {\bibinfo {title}
  {Selective linear or quadratic optomechanical coupling via measurement},}\
  }\href {\doibase 10.1103/PhysRevX.1.021011} {\bibfield  {journal} {\bibinfo
  {journal} {Physical Review X}\ }\textbf {\bibinfo {volume} {1}},\ \bibinfo
  {pages} {021011} (\bibinfo {year} {2011})}\BibitemShut {NoStop}%
\bibitem [{\citenamefont {Brawley}\ \emph {et~al.}(2016)\citenamefont
  {Brawley}, \citenamefont {Vanner}, \citenamefont {Larsen}, \citenamefont
  {Schmid}, \citenamefont {Boisen},\ and\ \citenamefont
  {Bowen}}]{brawley_nonlinear_2016}%
  \BibitemOpen
  \bibfield  {author} {\bibinfo {author} {\bibfnamefont {G.~A.}\ \bibnamefont
  {Brawley}}, \bibinfo {author} {\bibfnamefont {M.~R.}\ \bibnamefont {Vanner}},
  \bibinfo {author} {\bibfnamefont {P.~E.}\ \bibnamefont {Larsen}}, \bibinfo
  {author} {\bibfnamefont {S.}~\bibnamefont {Schmid}}, \bibinfo {author}
  {\bibfnamefont {A.}~\bibnamefont {Boisen}}, \ and\ \bibinfo {author}
  {\bibfnamefont {W.~P.}\ \bibnamefont {Bowen}},\ }\bibfield  {title} {\enquote
  {\bibinfo {title} {Nonlinear optomechanical measurement of mechanical
  motion},}\ }\href {\doibase 10.1038/ncomms10988} {\bibfield  {journal}
  {\bibinfo  {journal} {Nature Communications}\ }\textbf {\bibinfo {volume}
  {7}},\ \bibinfo {pages} {10988} (\bibinfo {year} {2016})},\ \bibinfo {note}
  {publisher: Nature Publishing Group}\BibitemShut {NoStop}%
\bibitem [{\citenamefont {Leijssen}\ \emph {et~al.}(2017)\citenamefont
  {Leijssen}, \citenamefont {Gala}, \citenamefont {Freisem}, \citenamefont
  {Muhonen},\ and\ \citenamefont {Verhagen}}]{leijssen_nonlinear_2017}%
  \BibitemOpen
  \bibfield  {author} {\bibinfo {author} {\bibfnamefont {Rick}\ \bibnamefont
  {Leijssen}}, \bibinfo {author} {\bibfnamefont {Giada R.~La}\ \bibnamefont
  {Gala}}, \bibinfo {author} {\bibfnamefont {Lars}\ \bibnamefont {Freisem}},
  \bibinfo {author} {\bibfnamefont {Juha~T.}\ \bibnamefont {Muhonen}}, \ and\
  \bibinfo {author} {\bibfnamefont {Ewold}\ \bibnamefont {Verhagen}},\
  }\bibfield  {title} {\enquote {\bibinfo {title} {Nonlinear cavity
  optomechanics with nanomechanical thermal fluctuations},}\ }\href {\doibase
  10.1038/ncomms16024} {\bibfield  {journal} {\bibinfo  {journal} {Nature
  Communications}\ }\textbf {\bibinfo {volume} {8}},\ \bibinfo {pages} {1--10}
  (\bibinfo {year} {2017})}\BibitemShut {NoStop}%
\bibitem [{\citenamefont {Thompson}\ \emph {et~al.}(2008)\citenamefont
  {Thompson}, \citenamefont {Zwickl}, \citenamefont {Jayich}, \citenamefont
  {Marquardt}, \citenamefont {Girvin},\ and\ \citenamefont
  {Harris}}]{thompson_strong_2008}%
  \BibitemOpen
  \bibfield  {author} {\bibinfo {author} {\bibfnamefont {J.~D.}\ \bibnamefont
  {Thompson}}, \bibinfo {author} {\bibfnamefont {B.~M.}\ \bibnamefont
  {Zwickl}}, \bibinfo {author} {\bibfnamefont {A.~M.}\ \bibnamefont {Jayich}},
  \bibinfo {author} {\bibfnamefont {Florian}\ \bibnamefont {Marquardt}},
  \bibinfo {author} {\bibfnamefont {S.~M.}\ \bibnamefont {Girvin}}, \ and\
  \bibinfo {author} {\bibfnamefont {J.~G.~E.}\ \bibnamefont {Harris}},\
  }\bibfield  {title} {\enquote {\bibinfo {title} {Strong dispersive coupling
  of a high-finesse cavity to a micromechanical membrane},}\ }\href {\doibase
  10.1038/nature06715} {\bibfield  {journal} {\bibinfo  {journal} {Nature}\
  }\textbf {\bibinfo {volume} {452}},\ \bibinfo {pages} {72–75} (\bibinfo
  {year} {2008})}\BibitemShut {NoStop}%
\bibitem [{\citenamefont {Wilson}\ \emph {et~al.}(2009)\citenamefont {Wilson},
  \citenamefont {Regal}, \citenamefont {Papp},\ and\ \citenamefont
  {Kimble}}]{wilson_cavity_2009}%
  \BibitemOpen
  \bibfield  {author} {\bibinfo {author} {\bibfnamefont {D.~J.}\ \bibnamefont
  {Wilson}}, \bibinfo {author} {\bibfnamefont {C.~A.}\ \bibnamefont {Regal}},
  \bibinfo {author} {\bibfnamefont {S.~B.}\ \bibnamefont {Papp}}, \ and\
  \bibinfo {author} {\bibfnamefont {H.~J.}\ \bibnamefont {Kimble}},\ }\bibfield
   {title} {\enquote {\bibinfo {title} {Cavity {Optomechanics} with
  {Stoichiometric} {SiN} {Films}},}\ }\href {\doibase
  10.1103/PhysRevLett.103.207204} {\bibfield  {journal} {\bibinfo  {journal}
  {Physical Review Letters}\ }\textbf {\bibinfo {volume} {103}},\ \bibinfo
  {pages} {207204} (\bibinfo {year} {2009})}\BibitemShut {NoStop}%
\bibitem [{\citenamefont {Cripe}\ \emph {et~al.}(2019)\citenamefont {Cripe},
  \citenamefont {Aggarwal}, \citenamefont {Lanza}, \citenamefont {Libson},
  \citenamefont {Singh}, \citenamefont {Heu}, \citenamefont {Follman},
  \citenamefont {Cole}, \citenamefont {Mavalvala},\ and\ \citenamefont
  {Corbitt}}]{cripe_measurement_2019}%
  \BibitemOpen
  \bibfield  {author} {\bibinfo {author} {\bibfnamefont {Jonathan}\
  \bibnamefont {Cripe}}, \bibinfo {author} {\bibfnamefont {Nancy}\ \bibnamefont
  {Aggarwal}}, \bibinfo {author} {\bibfnamefont {Robert}\ \bibnamefont
  {Lanza}}, \bibinfo {author} {\bibfnamefont {Adam}\ \bibnamefont {Libson}},
  \bibinfo {author} {\bibfnamefont {Robinjeet}\ \bibnamefont {Singh}}, \bibinfo
  {author} {\bibfnamefont {Paula}\ \bibnamefont {Heu}}, \bibinfo {author}
  {\bibfnamefont {David}\ \bibnamefont {Follman}}, \bibinfo {author}
  {\bibfnamefont {Garrett~D.}\ \bibnamefont {Cole}}, \bibinfo {author}
  {\bibfnamefont {Nergis}\ \bibnamefont {Mavalvala}}, \ and\ \bibinfo {author}
  {\bibfnamefont {Thomas}\ \bibnamefont {Corbitt}},\ }\bibfield  {title}
  {\enquote {\bibinfo {title} {Measurement of quantum back action in the audio
  band at room temperature},}\ }\href {\doibase 10.1038/s41586-019-1051-4}
  {\bibfield  {journal} {\bibinfo  {journal} {Nature}\ }\textbf {\bibinfo
  {volume} {568}},\ \bibinfo {pages} {364--367} (\bibinfo {year}
  {2019})}\BibitemShut {NoStop}%
\bibitem [{\citenamefont {Yap}\ \emph {et~al.}(2019)\citenamefont {Yap},
  \citenamefont {Cripe}, \citenamefont {Mansell}, \citenamefont {McRae},
  \citenamefont {Ward}, \citenamefont {Slagmolen}, \citenamefont {Heu},
  \citenamefont {Follman}, \citenamefont {Cole}, \citenamefont {Corbitt},\ and\
  \citenamefont {McClelland}}]{yap_broadband_2019}%
  \BibitemOpen
  \bibfield  {author} {\bibinfo {author} {\bibfnamefont {Min~Jet}\ \bibnamefont
  {Yap}}, \bibinfo {author} {\bibfnamefont {Jonathan}\ \bibnamefont {Cripe}},
  \bibinfo {author} {\bibfnamefont {Georgia~L.}\ \bibnamefont {Mansell}},
  \bibinfo {author} {\bibfnamefont {Terry~G.}\ \bibnamefont {McRae}}, \bibinfo
  {author} {\bibfnamefont {Robert~L.}\ \bibnamefont {Ward}}, \bibinfo {author}
  {\bibfnamefont {Bram J.~J.}\ \bibnamefont {Slagmolen}}, \bibinfo {author}
  {\bibfnamefont {Paula}\ \bibnamefont {Heu}}, \bibinfo {author} {\bibfnamefont
  {David}\ \bibnamefont {Follman}}, \bibinfo {author} {\bibfnamefont
  {Garrett~D.}\ \bibnamefont {Cole}}, \bibinfo {author} {\bibfnamefont
  {Thomas}\ \bibnamefont {Corbitt}}, \ and\ \bibinfo {author} {\bibfnamefont
  {David~E.}\ \bibnamefont {McClelland}},\ }\bibfield  {title} {\enquote
  {\bibinfo {title} {Broadband reduction of quantum radiation pressure noise
  via squeezed light injection},}\ }\href {\doibase 10.1038/s41566-019-0527-y}
  {\bibfield  {journal} {\bibinfo  {journal} {Nature Photonics}\ ,\ \bibinfo
  {pages} {1--5}} (\bibinfo {year} {2019})}\BibitemShut {NoStop}%
\bibitem [{\citenamefont {Tsaturyan}\ \emph {et~al.}(2017)\citenamefont
  {Tsaturyan}, \citenamefont {Barg}, \citenamefont {Polzik},\ and\
  \citenamefont {Schliesser}}]{tsaturyan_ultracoherent_2017}%
  \BibitemOpen
  \bibfield  {author} {\bibinfo {author} {\bibfnamefont {Y.}~\bibnamefont
  {Tsaturyan}}, \bibinfo {author} {\bibfnamefont {A.}~\bibnamefont {Barg}},
  \bibinfo {author} {\bibfnamefont {E.~S.}\ \bibnamefont {Polzik}}, \ and\
  \bibinfo {author} {\bibfnamefont {A.}~\bibnamefont {Schliesser}},\ }\bibfield
   {title} {\enquote {\bibinfo {title} {Ultracoherent nanomechanical resonators
  via soft clamping and dissipation dilution},}\ }\href {\doibase
  10.1038/nnano.2017.101} {\bibfield  {journal} {\bibinfo  {journal} {Nature
  Nanotechnology}\ }\textbf {\bibinfo {volume} {12}},\ \bibinfo {pages} {776}
  (\bibinfo {year} {2017})}\BibitemShut {NoStop}%
\bibitem [{\citenamefont {Reetz}\ \emph {et~al.}(2019)\citenamefont {Reetz},
  \citenamefont {Fischer}, \citenamefont {Assumpção}, \citenamefont
  {McNally}, \citenamefont {Burns}, \citenamefont {Sankey},\ and\ \citenamefont
  {Regal}}]{reetz_analysis_2019}%
  \BibitemOpen
  \bibfield  {author} {\bibinfo {author} {\bibfnamefont {C.}~\bibnamefont
  {Reetz}}, \bibinfo {author} {\bibfnamefont {R.}~\bibnamefont {Fischer}},
  \bibinfo {author} {\bibfnamefont {G.G.T.}\ \bibnamefont {Assumpção}},
  \bibinfo {author} {\bibfnamefont {D.P.}\ \bibnamefont {McNally}}, \bibinfo
  {author} {\bibfnamefont {P.S.}\ \bibnamefont {Burns}}, \bibinfo {author}
  {\bibfnamefont {J.C.}\ \bibnamefont {Sankey}}, \ and\ \bibinfo {author}
  {\bibfnamefont {C.A.}\ \bibnamefont {Regal}},\ }\bibfield  {title} {\enquote
  {\bibinfo {title} {Analysis of {Membrane} {Phononic} {Crystals} with {Wide}
  {Band} {Gaps} and {Low}-{Mass} {Defects}},}\ }\href {\doibase
  10.1103/PhysRevApplied.12.044027} {\bibfield  {journal} {\bibinfo  {journal}
  {Physical Review Applied}\ }\textbf {\bibinfo {volume} {12}},\ \bibinfo
  {pages} {044027} (\bibinfo {year} {2019})}\BibitemShut {NoStop}%
\bibitem [{\citenamefont {Vyatchanin}\ and\ \citenamefont
  {Zubova}(1995)}]{vyatchanin_quantum_1995}%
  \BibitemOpen
  \bibfield  {author} {\bibinfo {author} {\bibfnamefont {S.~P.}\ \bibnamefont
  {Vyatchanin}}\ and\ \bibinfo {author} {\bibfnamefont {E.~A.}\ \bibnamefont
  {Zubova}},\ }\bibfield  {title} {\enquote {\bibinfo {title} {Quantum
  variation measurement of a force},}\ }\href {\doibase
  10.1016/0375-9601(95)00280-G} {\bibfield  {journal} {\bibinfo  {journal}
  {Physics Letters A}\ }\textbf {\bibinfo {volume} {201}},\ \bibinfo {pages}
  {269--274} (\bibinfo {year} {1995})}\BibitemShut {NoStop}%
\bibitem [{\citenamefont {Sudhir}\ \emph
  {et~al.}(2017{\natexlab{a}})\citenamefont {Sudhir}, \citenamefont
  {Schilling}, \citenamefont {Fedorov}, \citenamefont {Schütz}, \citenamefont
  {Wilson},\ and\ \citenamefont {Kippenberg}}]{sudhir_quantum_2017}%
  \BibitemOpen
  \bibfield  {author} {\bibinfo {author} {\bibfnamefont {V.}~\bibnamefont
  {Sudhir}}, \bibinfo {author} {\bibfnamefont {R.}~\bibnamefont {Schilling}},
  \bibinfo {author} {\bibfnamefont {S. A.}\ \bibnamefont {Fedorov}}, \bibinfo
  {author} {\bibfnamefont {H.}~\bibnamefont {Schütz}}, \bibinfo {author}
  {\bibfnamefont {D. J.}\ \bibnamefont {Wilson}}, \ and\ \bibinfo {author}
  {\bibfnamefont {T. J.}\ \bibnamefont {Kippenberg}},\ }\bibfield  {title}
  {\enquote {\bibinfo {title} {Quantum {Correlations} of {Light} from a
  {Room}-{Temperature} {Mechanical} {Oscillator}},}\ }\href {\doibase
  10.1103/PhysRevX.7.031055} {\bibfield  {journal} {\bibinfo  {journal}
  {Physical Review X}\ }\textbf {\bibinfo {volume} {7}},\ \bibinfo {pages}
  {031055} (\bibinfo {year} {2017}{\natexlab{a}})}\BibitemShut {NoStop}%
\bibitem [{\citenamefont {Kampel}\ \emph {et~al.}(2017)\citenamefont {Kampel},
  \citenamefont {Peterson}, \citenamefont {Fischer}, \citenamefont {Yu},
  \citenamefont {Cicak}, \citenamefont {Simmonds}, \citenamefont {Lehnert},\
  and\ \citenamefont {Regal}}]{kampel_improving_2017}%
  \BibitemOpen
  \bibfield  {author} {\bibinfo {author} {\bibfnamefont {N. S.}\ \bibnamefont
  {Kampel}}, \bibinfo {author} {\bibfnamefont {R. W.}\ \bibnamefont
  {Peterson}}, \bibinfo {author} {\bibfnamefont {R.}~\bibnamefont {Fischer}},
  \bibinfo {author} {\bibfnamefont {P.-L.}\ \bibnamefont {Yu}}, \bibinfo
  {author} {\bibfnamefont {K.}~\bibnamefont {Cicak}}, \bibinfo {author}
  {\bibfnamefont {R. W.}\ \bibnamefont {Simmonds}}, \bibinfo {author}
  {\bibfnamefont {K. W.}\ \bibnamefont {Lehnert}}, \ and\ \bibinfo {author}
  {\bibfnamefont {C. A.}\ \bibnamefont {Regal}},\ }\bibfield  {title}
  {\enquote {\bibinfo {title} {Improving {Broadband} {Displacement} {Detection}
  with {Quantum} {Correlations}},}\ }\href {\doibase 10.1103/PhysRevX.7.021008}
  {\bibfield  {journal} {\bibinfo  {journal} {Physical Review X}\ }\textbf
  {\bibinfo {volume} {7}},\ \bibinfo {pages} {021008} (\bibinfo {year}
  {2017})}\BibitemShut {NoStop}%
\bibitem [{\citenamefont {Safavi-Naeini}\ \emph {et~al.}(2013)\citenamefont
  {Safavi-Naeini}, \citenamefont {Gröblacher}, \citenamefont {Hill},
  \citenamefont {Chan}, \citenamefont {Aspelmeyer},\ and\ \citenamefont
  {Painter}}]{safavi-naeini_squeezed_2013}%
  \BibitemOpen
  \bibfield  {author} {\bibinfo {author} {\bibfnamefont {Amir~H.}\ \bibnamefont
  {Safavi-Naeini}}, \bibinfo {author} {\bibfnamefont {Simon}\ \bibnamefont
  {Gröblacher}}, \bibinfo {author} {\bibfnamefont {Jeff~T.}\ \bibnamefont
  {Hill}}, \bibinfo {author} {\bibfnamefont {Jasper}\ \bibnamefont {Chan}},
  \bibinfo {author} {\bibfnamefont {Markus}\ \bibnamefont {Aspelmeyer}}, \ and\
  \bibinfo {author} {\bibfnamefont {Oskar}\ \bibnamefont {Painter}},\
  }\bibfield  {title} {\enquote {\bibinfo {title} {Squeezed light from a
  silicon micromechanical resonator},}\ }\href {\doibase 10.1038/nature12307}
  {\bibfield  {journal} {\bibinfo  {journal} {Nature}\ }\textbf {\bibinfo
  {volume} {500}},\ \bibinfo {pages} {185--189} (\bibinfo {year}
  {2013})}\BibitemShut {NoStop}%
\bibitem [{\citenamefont {Purdy}\ \emph {et~al.}(2013)\citenamefont {Purdy},
  \citenamefont {Yu}, \citenamefont {Peterson}, \citenamefont {Kampel},\ and\
  \citenamefont {Regal}}]{purdy_strong_2013}%
  \BibitemOpen
  \bibfield  {author} {\bibinfo {author} {\bibfnamefont {T.~P.}\ \bibnamefont
  {Purdy}}, \bibinfo {author} {\bibfnamefont {P.-L.}\ \bibnamefont {Yu}},
  \bibinfo {author} {\bibfnamefont {R.~W.}\ \bibnamefont {Peterson}}, \bibinfo
  {author} {\bibfnamefont {N.~S.}\ \bibnamefont {Kampel}}, \ and\ \bibinfo
  {author} {\bibfnamefont {C.~A.}\ \bibnamefont {Regal}},\ }\bibfield  {title}
  {\enquote {\bibinfo {title} {Strong {Optomechanical} {Squeezing} of
  {Light}},}\ }\href {\doibase 10.1103/PhysRevX.3.031012} {\bibfield  {journal}
  {\bibinfo  {journal} {Physical Review X}\ }\textbf {\bibinfo {volume} {3}},\
  \bibinfo {pages} {031012} (\bibinfo {year} {2013})}\BibitemShut {NoStop}%
\bibitem [{spe()}]{specta_notations}%
  \BibitemOpen
  \href@noop {} {}\bibinfo {howpublished} {We use two-sided spectral densities,
  denoted as $S_{xx}[\omega]$, in theoretical derivations and one-sided
  spectral densities, denoted as $S_{x}[\omega]=2S_{xx}[\omega]$ for
  $\omega>0$, for the presentation of experimental data.}\BibitemShut {Stop}%
\bibitem [{\citenamefont {Gardiner}()}]{gardiner_handbook_1985}%
  \BibitemOpen
  \bibfield  {author} {\bibinfo {author} {\bibfnamefont {Crispin~W.}\
  \bibnamefont {Gardiner}},\ }\href@noop {} {\emph {\bibinfo {title} {Handbook
  of Stochastic Methods}}},\ \bibinfo {edition} {2nd}\ ed.\ (\bibinfo
  {publisher} {Springer},\ \bibinfo {address} {Berlin})\ \bibinfo {note}
  {section 2.8.1}\BibitemShut {NoStop}%
\bibitem [{\citenamefont {Sudhir}\ \emph
  {et~al.}(2017{\natexlab{b}})\citenamefont {Sudhir}, \citenamefont {Wilson},
  \citenamefont {Schilling}, \citenamefont {Schütz}, \citenamefont {Fedorov},
  \citenamefont {Ghadimi}, \citenamefont {Nunnenkamp},\ and\ \citenamefont
  {Kippenberg}}]{sudhir_appearance_2017}%
  \BibitemOpen
  \bibfield  {author} {\bibinfo {author} {\bibfnamefont {V.}~\bibnamefont
  {Sudhir}}, \bibinfo {author} {\bibfnamefont {D. J.}\ \bibnamefont
  {Wilson}}, \bibinfo {author} {\bibfnamefont {R.}~\bibnamefont {Schilling}},
  \bibinfo {author} {\bibfnamefont {H.}~\bibnamefont {Schütz}}, \bibinfo
  {author} {\bibfnamefont {S. A.}\ \bibnamefont {Fedorov}}, \bibinfo {author}
  {\bibfnamefont {A. H.}\ \bibnamefont {Ghadimi}}, \bibinfo {author}
  {\bibfnamefont {A.}~\bibnamefont {Nunnenkamp}}, \ and\ \bibinfo {author}
  {\bibfnamefont {T. J.}\ \bibnamefont {Kippenberg}},\ }\bibfield  {title}
  {\enquote {\bibinfo {title} {Appearance and {Disappearance} of {Quantum}
  {Correlations} in {Measurement}-{Based} {Feedback} {Control} of a
  {Mechanical} {Oscillator}},}\ }\href {\doibase 10.1103/PhysRevX.7.011001}
  {\bibfield  {journal} {\bibinfo  {journal} {Physical Review X}\ }\textbf
  {\bibinfo {volume} {7}},\ \bibinfo {pages} {011001} (\bibinfo {year}
  {2017}{\natexlab{b}})}\BibitemShut {NoStop}%
\bibitem [{\citenamefont {Zhao}\ \emph {et~al.}(2012)\citenamefont {Zhao},
  \citenamefont {Wilson}, \citenamefont {Ni},\ and\ \citenamefont
  {Kimble}}]{zhao_wilson_suppression_2012}%
  \BibitemOpen
  \bibfield  {author} {\bibinfo {author} {\bibfnamefont {Yi}~\bibnamefont
  {Zhao}}, \bibinfo {author} {\bibfnamefont {Dalziel~J.}\ \bibnamefont
  {Wilson}}, \bibinfo {author} {\bibfnamefont {K.-K.}\ \bibnamefont {Ni}}, \
  and\ \bibinfo {author} {\bibfnamefont {H.~J.}\ \bibnamefont {Kimble}},\
  }\bibfield  {title} {\enquote {\bibinfo {title} {Suppression of extraneous
  thermal noise in cavity optomechanics},}\ }\href {\doibase
  10.1364/OE.20.003586} {\bibfield  {journal} {\bibinfo  {journal} {Optics
  Express}\ }\textbf {\bibinfo {volume} {20}},\ \bibinfo {pages} {3586--3612}
  (\bibinfo {year} {2012})},\ \bibinfo {note} {publisher: Optical Society of
  America}\BibitemShut {NoStop}%
\bibitem [{\citenamefont {Wilson}(2012)}]{wilson_thesis_2012}%
  \BibitemOpen
  \bibfield  {author} {\bibinfo {author} {\bibfnamefont {Dalziel~Joseph}\
  \bibnamefont {Wilson}},\ }\emph {\bibinfo {title} {Cavity optomechanics with
  high-stress silicon nitride films}},\ \href
  {http://resolver.caltech.edu/CaltechTHESIS:06122012-123343193} {\bibinfo
  {type} {Phd thesis}},\ \bibinfo  {school} {California Institute of
  Technology} (\bibinfo {year} {2012})\BibitemShut {NoStop}%
\bibitem [{\citenamefont {Ghadimi}\ \emph {et~al.}(2018)\citenamefont
  {Ghadimi}, \citenamefont {Fedorov}, \citenamefont {Engelsen}, \citenamefont
  {Bereyhi}, \citenamefont {Schilling}, \citenamefont {Wilson},\ and\
  \citenamefont {Kippenberg}}]{ghadimi_strain_2017}%
  \BibitemOpen
  \bibfield  {author} {\bibinfo {author} {\bibfnamefont {A.~H.}\ \bibnamefont
  {Ghadimi}}, \bibinfo {author} {\bibfnamefont {S.~A.}\ \bibnamefont
  {Fedorov}}, \bibinfo {author} {\bibfnamefont {N.~J.}\ \bibnamefont
  {Engelsen}}, \bibinfo {author} {\bibfnamefont {M.~J.}\ \bibnamefont
  {Bereyhi}}, \bibinfo {author} {\bibfnamefont {R.}~\bibnamefont {Schilling}},
  \bibinfo {author} {\bibfnamefont {D.~J.}\ \bibnamefont {Wilson}}, \ and\
  \bibinfo {author} {\bibfnamefont {T.~J.}\ \bibnamefont {Kippenberg}},\
  }\bibfield  {title} {\enquote {\bibinfo {title} {Elastic strain engineering
  for ultralow mechanical dissipation},}\ }\href {\doibase
  10.1126/science.aar6939} {\bibfield  {journal} {\bibinfo  {journal}
  {Science}\ }\textbf {\bibinfo {volume} {360}},\ \bibinfo {pages} {764--768}
  (\bibinfo {year} {2018})}\BibitemShut {NoStop}%
\bibitem [{\citenamefont {Rossi}\ \emph {et~al.}(2018)\citenamefont {Rossi},
  \citenamefont {Mason}, \citenamefont {Chen}, \citenamefont {Tsaturyan},\ and\
  \citenamefont {Schliesser}}]{rossi_measurement-based_2018}%
  \BibitemOpen
  \bibfield  {author} {\bibinfo {author} {\bibfnamefont {Massimiliano}\
  \bibnamefont {Rossi}}, \bibinfo {author} {\bibfnamefont {David}\ \bibnamefont
  {Mason}}, \bibinfo {author} {\bibfnamefont {Junxin}\ \bibnamefont {Chen}},
  \bibinfo {author} {\bibfnamefont {Yeghishe}\ \bibnamefont {Tsaturyan}}, \
  and\ \bibinfo {author} {\bibfnamefont {Albert}\ \bibnamefont {Schliesser}},\
  }\bibfield  {title} {\enquote {\bibinfo {title} {Measurement-based quantum
  control of mechanical motion},}\ }\href {\doibase 10.1038/s41586-018-0643-8}
  {\bibfield  {journal} {\bibinfo  {journal} {Nature}\ }\textbf {\bibinfo
  {volume} {563}},\ \bibinfo {pages} {53} (\bibinfo {year} {2018})}\BibitemShut
  {NoStop}%
\bibitem [{\citenamefont {Martin}\ and\ \citenamefont
  {Zurek}(2007)}]{martin_measurement_2007}%
  \BibitemOpen
  \bibfield  {author} {\bibinfo {author} {\bibfnamefont {I.}~\bibnamefont
  {Martin}}\ and\ \bibinfo {author} {\bibfnamefont {W.~H.}\ \bibnamefont
  {Zurek}},\ }\bibfield  {title} {\enquote {\bibinfo {title} {Measurement of
  energy eigenstates by a slow detector},}\ }\href {\doibase
  10.1103/PhysRevLett.98.120401} {\bibfield  {journal} {\bibinfo  {journal}
  {Physical Review Letters}\ }\textbf {\bibinfo {volume} {98}},\ \bibinfo
  {pages} {120401} (\bibinfo {year} {2007})}\BibitemShut {NoStop}%
\bibitem [{\citenamefont {Gangat}\ \emph {et~al.}(2011)\citenamefont {Gangat},
  \citenamefont {Stace},\ and\ \citenamefont {Milburn}}]{gangat_phonon_2011}%
  \BibitemOpen
  \bibfield  {author} {\bibinfo {author} {\bibfnamefont {A.~A.}\ \bibnamefont
  {Gangat}}, \bibinfo {author} {\bibfnamefont {T.~M.}\ \bibnamefont {Stace}}, \
  and\ \bibinfo {author} {\bibfnamefont {G.~J.}\ \bibnamefont {Milburn}},\
  }\bibfield  {title} {\enquote {\bibinfo {title} {Phonon number quantum jumps
  in an optomechanical system},}\ }\href {\doibase
  10.1088/1367-2630/13/4/043024} {\bibfield  {journal} {\bibinfo  {journal}
  {New Journal of Physics}\ }\textbf {\bibinfo {volume} {13}},\ \bibinfo
  {pages} {043024} (\bibinfo {year} {2011})}\BibitemShut {NoStop}%
\bibitem [{\citenamefont {Braginsky}\ \emph {et~al.}(1992)\citenamefont
  {Braginsky}, \citenamefont {Khalili},\ and\ \citenamefont
  {Thorne}}]{braginsky_quantum_1992}%
  \BibitemOpen
  \bibfield  {author} {\bibinfo {author} {\bibfnamefont {Vladimir~B.}\
  \bibnamefont {Braginsky}}, \bibinfo {author} {\bibfnamefont {Farid~Ya}\
  \bibnamefont {Khalili}}, \ and\ \bibinfo {author} {\bibfnamefont {Kip~S.}\
  \bibnamefont {Thorne}},\ }\href@noop {} {\emph {\bibinfo {title} {Quantum
  {Measurement}}}},\ \bibinfo {edition} {1st}\ ed.\ (\bibinfo  {publisher}
  {Cambridge University Press},\ \bibinfo {address} {Cambridge England ; New
  York, NY, USA},\ \bibinfo {year} {1992})\BibitemShut {NoStop}%
\bibitem [{\citenamefont {Clerk}\ \emph
  {et~al.}(2010{\natexlab{b}})\citenamefont {Clerk}, \citenamefont
  {Marquardt},\ and\ \citenamefont {Harris}}]{clerk_quantum_2010}%
  \BibitemOpen
  \bibfield  {author} {\bibinfo {author} {\bibfnamefont {A.~A.}\ \bibnamefont
  {Clerk}}, \bibinfo {author} {\bibfnamefont {Florian}\ \bibnamefont
  {Marquardt}}, \ and\ \bibinfo {author} {\bibfnamefont {J.~G.~E.}\
  \bibnamefont {Harris}},\ }\bibfield  {title} {\enquote {\bibinfo {title}
  {Quantum {Measurement} of {Phonon} {Shot} {Noise}},}\ }\href {\doibase
  10.1103/PhysRevLett.104.213603} {\bibfield  {journal} {\bibinfo  {journal}
  {Physical Review Letters}\ }\textbf {\bibinfo {volume} {104}},\ \bibinfo
  {pages} {213603} (\bibinfo {year} {2010}{\natexlab{b}})}\BibitemShut
  {NoStop}%
\bibitem [{\citenamefont {Nunnenkamp}\ \emph {et~al.}(2010)\citenamefont
  {Nunnenkamp}, \citenamefont {Børkje}, \citenamefont {Harris},\ and\
  \citenamefont {Girvin}}]{nunnenkamp_cooling_and_squeezing_2010}%
  \BibitemOpen
  \bibfield  {author} {\bibinfo {author} {\bibfnamefont {A.}~\bibnamefont
  {Nunnenkamp}}, \bibinfo {author} {\bibfnamefont {K.}~\bibnamefont {Børkje}},
  \bibinfo {author} {\bibfnamefont {J.~G.~E.}\ \bibnamefont {Harris}}, \ and\
  \bibinfo {author} {\bibfnamefont {S.~M.}\ \bibnamefont {Girvin}},\ }\bibfield
   {title} {\enquote {\bibinfo {title} {Cooling and squeezing via quadratic
  optomechanical coupling},}\ }\href {\doibase 10.1103/PhysRevA.82.021806}
  {\bibfield  {journal} {\bibinfo  {journal} {Physical Review A}\ }\textbf
  {\bibinfo {volume} {82}},\ \bibinfo {pages} {021806} (\bibinfo {year}
  {2010})}\BibitemShut {NoStop}%
\bibitem [{\citenamefont {Paraïso}\ \emph {et~al.}(2015)\citenamefont
  {Paraïso}, \citenamefont {Kalaee}, \citenamefont {Zang}, \citenamefont
  {Pfeifer}, \citenamefont {Marquardt},\ and\ \citenamefont
  {Painter}}]{paraiso_position-squared_2015}%
  \BibitemOpen
  \bibfield  {author} {\bibinfo {author} {\bibfnamefont {Taofiq~K.}\
  \bibnamefont {Paraïso}}, \bibinfo {author} {\bibfnamefont {Mahmoud}\
  \bibnamefont {Kalaee}}, \bibinfo {author} {\bibfnamefont {Leyun}\
  \bibnamefont {Zang}}, \bibinfo {author} {\bibfnamefont {Hannes}\ \bibnamefont
  {Pfeifer}}, \bibinfo {author} {\bibfnamefont {Florian}\ \bibnamefont
  {Marquardt}}, \ and\ \bibinfo {author} {\bibfnamefont {Oskar}\ \bibnamefont
  {Painter}},\ }\bibfield  {title} {\enquote {\bibinfo {title}
  {Position-{Squared} {Coupling} in a {Tunable} {Photonic} {Crystal}
  {Optomechanical} {Cavity}},}\ }\href {\doibase 10.1103/PhysRevX.5.041024}
  {\bibfield  {journal} {\bibinfo  {journal} {Physical Review X}\ }\textbf
  {\bibinfo {volume} {5}},\ \bibinfo {pages} {041024} (\bibinfo {year}
  {2015})}\BibitemShut {NoStop}%
\bibitem [{\citenamefont {Clerk}(2004)}]{clerk_quantum-limited_2004}%
  \BibitemOpen
  \bibfield  {author} {\bibinfo {author} {\bibfnamefont {A.~A.}\ \bibnamefont
  {Clerk}},\ }\bibfield  {title} {\enquote {\bibinfo {title} {Quantum-limited
  position detection and amplification: A linear response perspective},}\
  }\href {\doibase 10.1103/PhysRevB.70.245306} {\bibfield  {journal} {\bibinfo
  {journal} {Physical Review B}\ }\textbf {\bibinfo {volume} {70}},\ \bibinfo
  {pages} {245306} (\bibinfo {year} {2004})},\ \bibinfo {note} {publisher:
  American Physical Society}\BibitemShut {NoStop}%
\bibitem [{\citenamefont {Lemonde}\ \emph {et~al.}(2013)\citenamefont
  {Lemonde}, \citenamefont {Didier},\ and\ \citenamefont
  {Clerk}}]{lemonde_nonlinear_2013}%
  \BibitemOpen
  \bibfield  {author} {\bibinfo {author} {\bibfnamefont {Marc-Antoine}\
  \bibnamefont {Lemonde}}, \bibinfo {author} {\bibfnamefont {Nicolas}\
  \bibnamefont {Didier}}, \ and\ \bibinfo {author} {\bibfnamefont {Aashish~A.}\
  \bibnamefont {Clerk}},\ }\bibfield  {title} {\enquote {\bibinfo {title}
  {Nonlinear interaction effects in a strongly driven optomechanical cavity},}\
  }\href {\doibase 10.1103/PhysRevLett.111.053602} {\bibfield  {journal}
  {\bibinfo  {journal} {Physical Review Letters}\ }\textbf {\bibinfo {volume}
  {111}},\ \bibinfo {pages} {053602} (\bibinfo {year} {2013})}\BibitemShut
  {NoStop}%
\bibitem [{\citenamefont {Matsko}\ and\ \citenamefont
  {Vyatchanin}(2018)}]{matsko_electromagnetic-continuum-induced_2018}%
  \BibitemOpen
  \bibfield  {author} {\bibinfo {author} {\bibfnamefont {Andrey~B.}\
  \bibnamefont {Matsko}}\ and\ \bibinfo {author} {\bibfnamefont {Sergey~P.}\
  \bibnamefont {Vyatchanin}},\ }\bibfield  {title} {\enquote {\bibinfo {title}
  {Electromagnetic-continuum-induced nonlinearity},}\ }\href {\doibase
  10.1103/PhysRevA.97.053824} {\bibfield  {journal} {\bibinfo  {journal}
  {Physical Review A}\ }\textbf {\bibinfo {volume} {97}},\ \bibinfo {pages}
  {053824} (\bibinfo {year} {2018})}\BibitemShut {NoStop}%
\bibitem [{\citenamefont {Reinhardt}\ \emph {et~al.}(2016)\citenamefont
  {Reinhardt}, \citenamefont {Müller}, \citenamefont {Bourassa},\ and\
  \citenamefont {Sankey}}]{reinhardt_ultralow-noise_2016}%
  \BibitemOpen
  \bibfield  {author} {\bibinfo {author} {\bibfnamefont {Christoph}\
  \bibnamefont {Reinhardt}}, \bibinfo {author} {\bibfnamefont {Tina}\
  \bibnamefont {Müller}}, \bibinfo {author} {\bibfnamefont {Alexandre}\
  \bibnamefont {Bourassa}}, \ and\ \bibinfo {author} {\bibfnamefont {Jack~C.}\
  \bibnamefont {Sankey}},\ }\bibfield  {title} {\enquote {\bibinfo {title}
  {Ultralow-{Noise} {SiN} {Trampoline} {Resonators} for {Sensing} and
  {Optomechanics}},}\ }\href {\doibase 10.1103/PhysRevX.6.021001} {\bibfield
  {journal} {\bibinfo  {journal} {Physical Review X}\ }\textbf {\bibinfo
  {volume} {6}},\ \bibinfo {pages} {021001} (\bibinfo {year}
  {2016})}\BibitemShut {NoStop}%
\bibitem [{\citenamefont {Fedorov}\ \emph {et~al.}(2020)\citenamefont
  {Fedorov}, \citenamefont {Beccari}, \citenamefont {Engelsen},\ and\
  \citenamefont {Kippenberg}}]{fedorov_fractal-like_2020}%
  \BibitemOpen
  \bibfield  {author} {\bibinfo {author} {\bibfnamefont {S. A.}\ \bibnamefont
  {Fedorov}}, \bibinfo {author} {\bibfnamefont {A.}~\bibnamefont {Beccari}},
  \bibinfo {author} {\bibfnamefont {N. J.}\ \bibnamefont {Engelsen}}, \ and\
  \bibinfo {author} {\bibfnamefont {T. J.}\ \bibnamefont {Kippenberg}},\
  }\bibfield  {title} {\enquote {\bibinfo {title} {Fractal-like {Mechanical}
  {Resonators} with a {Soft}-{Clamped} {Fundamental} {Mode}},}\ }\href
  {\doibase 10.1103/PhysRevLett.124.025502} {\bibfield  {journal} {\bibinfo
  {journal} {Physical Review Letters}\ }\textbf {\bibinfo {volume} {124}},\
  \bibinfo {pages} {025502} (\bibinfo {year} {2020})}\BibitemShut {NoStop}%
\bibitem [{zen()}]{zenodo_repos}%
  \BibitemOpen
  \href@noop {} {}\bibinfo {howpublished} {Raw measurements data, analysis code
  to reproduce the manuscript figures, and the GDS designs of PnC membranes are
  available on zenodo.org, DOI: 10.5281/zenodo.3747301}\BibitemShut {NoStop}%
\bibitem [{\citenamefont {Gorodetksy}\ \emph {et~al.}(2010)\citenamefont
  {Gorodetksy}, \citenamefont {Schliesser}, \citenamefont {Anetsberger},
  \citenamefont {Deleglise},\ and\ \citenamefont
  {Kippenberg}}]{gorodetksy_determination_2010}%
  \BibitemOpen
  \bibfield  {author} {\bibinfo {author} {\bibfnamefont {M.~L.}\ \bibnamefont
  {Gorodetksy}}, \bibinfo {author} {\bibfnamefont {A.}~\bibnamefont
  {Schliesser}}, \bibinfo {author} {\bibfnamefont {G.}~\bibnamefont
  {Anetsberger}}, \bibinfo {author} {\bibfnamefont {S.}~\bibnamefont
  {Deleglise}}, \ and\ \bibinfo {author} {\bibfnamefont {T.~J.}\ \bibnamefont
  {Kippenberg}},\ }\bibfield  {title} {\enquote {\bibinfo {title}
  {Determination of the vacuum optomechanical coupling rate using frequency
  noise calibration},}\ }\href {\doibase 10.1364/OE.18.023236} {\bibfield
  {journal} {\bibinfo  {journal} {Optics Express}\ }\textbf {\bibinfo {volume}
  {18}},\ \bibinfo {pages} {23236--23246} (\bibinfo {year} {2010})}\BibitemShut
  {NoStop}%
\bibitem [{\citenamefont {Gärtner}\ \emph {et~al.}(2018)\citenamefont
  {Gärtner}, \citenamefont {Moura}, \citenamefont {Haaxman}, \citenamefont
  {Norte},\ and\ \citenamefont {Gröblacher}}]{gartner2018integrated}%
  \BibitemOpen
  \bibfield  {author} {\bibinfo {author} {\bibfnamefont {Claus}\ \bibnamefont
  {Gärtner}}, \bibinfo {author} {\bibfnamefont {Jo{\~a}o~P}\ \bibnamefont
  {Moura}}, \bibinfo {author} {\bibfnamefont {Wouter}\ \bibnamefont {Haaxman}},
  \bibinfo {author} {\bibfnamefont {Richard~A}\ \bibnamefont {Norte}}, \ and\
  \bibinfo {author} {\bibfnamefont {Simon}\ \bibnamefont {Gröblacher}},\
  }\bibfield  {title} {\enquote {\bibinfo {title} {Integrated optomechanical
  arrays of two high reflectivity sin membranes},}\ }\href@noop {} {\bibfield
  {journal} {\bibinfo  {journal} {Nano Letters}\ }\textbf {\bibinfo {volume}
  {18}},\ \bibinfo {pages} {7171--7175} (\bibinfo {year} {2018})}\BibitemShut
  {NoStop}%
\bibitem [{\citenamefont {Nielsen}\ \emph {et~al.}(2004)\citenamefont
  {Nielsen}, \citenamefont {Christensen}, \citenamefont {Pedersen},\ and\
  \citenamefont {Thomsen}}]{nielsen2004particle}%
  \BibitemOpen
  \bibfield  {author} {\bibinfo {author} {\bibfnamefont {C~Bergenstof}\
  \bibnamefont {Nielsen}}, \bibinfo {author} {\bibfnamefont {Carsten}\
  \bibnamefont {Christensen}}, \bibinfo {author} {\bibfnamefont {Casper}\
  \bibnamefont {Pedersen}}, \ and\ \bibinfo {author} {\bibfnamefont
  {Erik~Vilain}\ \bibnamefont {Thomsen}},\ }\bibfield  {title} {\enquote
  {\bibinfo {title} {Particle precipitation in connection with koh etching of
  silicon},}\ }\href@noop {} {\bibfield  {journal} {\bibinfo  {journal}
  {Journal of The Electrochemical Society}\ }\textbf {\bibinfo {volume}
  {151}},\ \bibinfo {pages} {G338--G342} (\bibinfo {year} {2004})}\BibitemShut
  {NoStop}%
\end{thebibliography}%

\end{document}